\documentclass[pdflatex,sn-aps]{sn-jnl}

\usepackage{graphicx}%
\usepackage{multirow}%
\usepackage{amsmath,amssymb,amsfonts}%
\usepackage{amsthm}%
\usepackage[title]{appendix}%
\usepackage{xcolor}%
\usepackage{textcomp}%
\usepackage{manyfoot}%
\usepackage{booktabs}%
\usepackage{algorithm}%
\usepackage{algorithmicx}%
\usepackage{algpseudocode}%
\usepackage{listings}%

\usepackage[T1]{fontenc}
\usepackage{cleveref}
\usepackage{tikz}
\usetikzlibrary{shapes, arrows.meta}
\usetikzlibrary{quantikz2}
\usepackage{subcaption}
\DeclareCaptionFont{white}{\color{white}}
\DeclareCaptionFormat{listing}{\colorbox{gray}{\parbox{\textwidth}{#1#2#3}}}
\lstdefinestyle{framed}
{
     frame=lrb,         
     belowcaptionskip=-1pt,
     xleftmargin=8pt,
     framexleftmargin=8pt,
     framexrightmargin=0pt,
     framextopmargin=5pt,
     framexbottommargin=5pt,
     framesep=0pt,
     rulesep=0pt,
 }
\DeclareCaptionFont{white}{\color{white}} 
\DeclareCaptionFormat{listing}{%
  \hspace*{-0.4pt}\colorbox{gray}{\parbox{\dimexpr\textwidth-2\fboxsep+.8pt\relax}{#1#2#3}}} 
\makeindex             
\makeatletter
\DeclareRobustCommand{\rvdots}{%
  \vbox{
    \baselineskip4\p@\lineskiplimit\z@
    \kern-\p@
    \hbox{.}\hbox{.}\hbox{.}
  }}
\makeatother

\theoremstyle{thmstyleone}%
\theoremstyle{thmstyletwo}%

\theoremstyle{thmstylethree}%

\begin{document}

\title[Quantum Optimization Algorithms]{Quantum Optimization Algorithms}


\author*{\fnm{Jonas} \sur{Stein}}\email{jonas.stein@ifi.lmu.de}

\author{\fnm{Maximilian} \sur{Zorn}}

\author{\fnm{Leo} \sur{Sünkel}}

\author{\fnm{Thomas} \sur{Gabor}}

\affil{\orgdiv{Institut für Informatik}, \orgname{LMU Munich}, \orgaddress{\street{Oettingenstraße 67}, \city{Munich}, \postcode{80538}, \state{Bavaria}, \country{Germany}}}


\abstract{Quantum optimization allows for up to exponential quantum speedups for specific, possibly industrially relevant problems. As the key algorithm in this field, we motivate and discuss the Quantum Approximate Optimization Algorithm (QAOA), which can be understood as a slightly generalized version of Quantum Annealing for gate-based quantum computers. We delve into the quantum circuit implementation of the QAOA, including Hamiltonian simulation techniques for higher-order Ising models, and discuss parameter training using the parameter shift rule. An example implementation with Pennylane source code demonstrates practical application for the Maximum Cut problem. Further, we show how constraints can be incorporated into the QAOA using Grover mixers, allowing to restrict the search space to strictly valid solutions for specific problems. Finally, we outline the Variational Quantum Eigensolver (VQE) as a generalization of the QAOA, highlighting its potential in the NISQ era and addressing challenges such as barren plateaus and ansatz design.}




\maketitle

\section{Employing Quantum-Mechanical Effects for Optimization}
Combinatorial optimization problems, such as determining the most efficient routing of delivery trucks or finding the optimal schedule for a set of tasks, involve searching through a vast number of possible solutions to find the best one. For a small subset of potentially practically relevant problems ranging from the most basic maximum cut problem~\cite{Tate2023warmstartedqaoa} to the optimization of higher order objective functions~\cite{PRXQuantum.5.030348,doi:10.1126/sciadv.adm6761}), there already exists promising evidence that the standard quantum optimization algorithm for gate model quantum computer, called Quantum Approximate Optimization Algorithm (QAOA), can outperform classical state-of-the-art solvers. Importantly, it is also known that the QAOA cannot be simulated efficiently using purely classical compute without overthrowing existing complexity classes separating quantum from classical computation efficiency~\cite{farhi2019quantumsupremacyquantumapproximate}. Quantum computing hence offers promising approaches to tackle these complex problems by exploiting quantum mechanical principles. A fundamental quantum principle used in Quantum Annealing as well as  the QAOA is the \textbf{adiabatic theorem}.

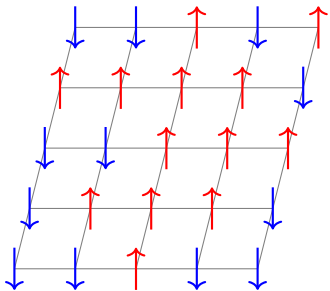
\begin{figure}[hbtp]
\centering
\begin{tikzpicture}[scale=0.8]
    \def\rows{3}  
    \def\cols{4}  

     \foreach \x in {0,...,\cols} {
        \draw[gray, thin] (\x * 0.25, \x) -- (\x * 0.25 + \cols, \x );
        \draw[gray, thin] (\x , 0) -- (\x + 1, \rows + 1);
    }

    \foreach \x in {0,...,\cols} {
        \foreach \y in {0,...,4} {
            \pgfmathtruncatemacro{\spin}{ifthenelse(int(random()*2) == 0, 1, -1)}
            \ifnum\spin=1
                \draw[thick, red, ->] (\x+ \y * 0.25, \y - 0.35) -- (\x+ \y * 0.25, \y + 0.35);
            \else
                \draw[thick, blue, ->] (\x + \y * 0.25, \y + 0.35) -- (\x+ \y * 0.25, \y - 0.35);
            \fi
        }
    }
\end{tikzpicture}
\caption{Standard grid-based form of a classical Ising model with spins pointing up or down.}
\label{fig:ising}
\end{figure}

To understand how the adiabatic theorem applies to solving combinatorial optimization problems, it is helpful to revisit the \textbf{Ising model}---a foundational concept in statistical physics that serves as a bridge between optimization problems and quantum mechanics. One classical instance of an Ising model can be visualized, e.g., as a grid or lattice of tiny bar magnets as shown in \Cref{fig:ising}, each representing a ``spin'' that can point up ($+1$) or down ($-1$). These spins interact with their nearest neighbors magnetically, and each spin can also be influenced by an external magnetic field. The interactions between neighboring spins are characterized by coupling constants $J_{ij}\in\mathbb{R}$, which determine whether adjacent spins prefer to align in the same direction (ferromagnetic interaction) or in opposite directions (antiferromagnetic interaction). Additionally, each spin can be subjected to its own external magnetic field $h_i\in\mathbb{R}$, allowing for individual biases.

In this model, the spins represent magnetic moments, analogous to tiny compass needles that can align parallel or anti-parallel to an applied external magnetic field. The energy of a particular spin configuration in the Ising model is given by the \textbf{classical Ising Hamiltonian}:
\begin{equation}
    E = -\sum_{\langle i,j \rangle} J_{ij} s_i s_j - \sum_i h_i s_i,
\end{equation}
where $s_i = \pm 1$ is the spin at the site $i$, the first sum runs over pairs of neighboring spins denoted as $\langle i,j \rangle$, $J_{ij}$ is the interaction strength between spins $s_i$ and $s_j$, and $h_i$ is the external magnetic field acting on spin $s_i$.

Finding the spin configuration that minimizes this energy corresponds to finding the so-called \textbf{ground state} of the system. This task is analogous to solving a combinatorial optimization problem, as each possible configuration of spins represents a potential solution with respect to the Ising Hamiltonian as the objective function. Crucially, nearly all practically significant combinatorial optimization problems can be translated into the Ising model~\cite{lucas2014ising}. This mapping is achieved by formulating the optimization problem as an instance of the NP-hard \textbf{Quadratic Unconstrained Binary Optimization (QUBO)} problem, for which the objective is to minimize a quadratic function of binary variables. By associating the binary variables with the spin values $s_i$, the QUBO problem becomes equivalent to finding the ground state of the Ising Hamiltonian. This isomorphism allows us to use the Ising model as a framework to solve combinatorial optimization problems.

The \textbf{adiabatic theorem} becomes relevant when we consider how to evolve a quantum system from an initial state to the desired final state that encodes the solution to our optimization problem. The adiabatic theorem essentially states that if a quantum system is initialized in the ground state of some Hamiltonian $H(0)$ and if the Hamiltonian is changed slowly enough to a final Hamiltonian $H(T)$ over time $t\in[0,T]$, the system will remain in the ground state of the instantaneous Hamiltonian $H(t)$ at each moment, provided there are no energy level crossings~\cite{Born1928}.

Mathematically, the time evolution of the quantum state is described by the unitary operator:
\begin{equation}
U(T) = \exp\left( -i \int_0^T H(t) \, dt \right),    
\end{equation}
where $H(t)$ is the time-dependent Hamiltonian that smoothly interpolates between $H(0)$ and $H(T)$. The exponential of the integral of the Hamiltonian encapsulates the continuous transformation of the system's state due to the changing Hamiltonian. Note that time ordering is implicit in this expression, as the Hamiltonians at different times generally do not commute (cf. $e^{A+B}\neq e^A e^B$ if $AB\neq BA$).

In the context of \textbf{adiabatic quantum computing (AQC)}, we exploit the adiabatic theorem to find the ground state of a problem Hamiltonian $H_{\text{problem}}$ that encodes the cost landscape of our combinatorial optimization problem. The process involves the following steps:

\begin{enumerate}
    \item \textit{Initialization}: Prepare the quantum system in the ground state of an initial Hamiltonian $H(0) = H_{\text{initial}}$, which is selected to have a simple form and whose ground state is easy to prepare.
    \item \textit{Adiabatic Evolution}: Slowly evolve the Hamiltonian from $H_{\text{initial}}$ to the problem Hamiltonian $H_{\text{problem}}=H(T)$ over a total time $T$, according to a schedule defined by a monotonic function $s:[0,T]\rightarrow [0,1]$ that continuously interpolates from 0 to 1, ensuring a gradual transition:
    \begin{equation}
        H(t) = (1 - s(t)) H_{\text{initial}} + s(t) H_{\text{problem}}.
    \end{equation}
   \item \textit{Final State}: If the evolution is carried out slowly enough to satisfy the adiabatic condition, the system remains in the ground state throughout the process and ends up in the ground state of $H_{\text{problem}}$ at time $T$.
   \item \textit{Measurement}: Measure the final state of the system yielding the solution to the original optimization problem.
\end{enumerate}
The \textbf{adiabatic condition} requires that the rate of change of the Hamiltonian is much less than $1/g^2$, where $g=\min_t E_1(t) - E_0(t)$ denotes the smallest energy gap between the ground state and the first excited state, with $|\psi_0(t)\rangle$ and $|\psi_1(t)\rangle$ denoting the instantaneous ground and first excited states of $H(t)$, and $E_0(t)$ and $E_1(t)$ the corresponding energy levels.

In simple terms, this condition ensures that the probability of the system transitioning from the ground state to an excited state during the evolution is negligible. However, for NP-hard optimization problems, the \textbf{spectral gap} $g$ often becomes \textbf{exponentially small} as the size of the problem increases. This means that, to satisfy the adiabatic condition, the total evolution time $T$ can be exponentially long, indicating that this approach also suffers from exponential runtimes for solving NP-hard problems. However, up to exponential quantum speedups compared to brute-force search are still possible and have already been observed empirically for specific optimization problems~\cite{10313741}.

To allow for an implementation of such an adiabatic time evolution, we introduce the \textbf{transverse-field Ising model}, which is a special quantum version of the Ising model. In the generalization of the classical Ising model to the quantum Ising model, the variables $s_i = \pm 1$ are replaced by the Pauli-Z matrices $\sigma_i^z$:
\begin{equation}
H_{\text{problem}} = -\sum_{\langle i,j \rangle} J_{ij} \sigma_i^z \sigma_j^z - \sum_i h_i \sigma_i^z.    
\end{equation}
Therefore, this model is clearly still isomorphic to QUBO. In this context, the transverse-field Ising model allows us to start in an easy-to-prepare ground state of an initial Hamiltonian $H_{\text{initial}}$ term in the $x$-dimension:
\begin{equation}
H_{\text{initial}} = -\Gamma \sum_i \sigma_i^x, 
\end{equation}
where $\sigma_i^x$ are the Pauli-X matrices acting on spin $i$, and $\Gamma\in\mathbb{R}$ is the strength of the transverse magnetic field. The full transverse-field Ising model then takes the form of 
\begin{equation}
    H = -\sum_{\langle i,j \rangle} J_{ij} \sigma_i^z \sigma_j^z - \sum_i h_i \sigma_i^z -\Gamma \sum_i \sigma_i^x.
\end{equation}
This construction is particularly useful, as it allows us to represent QUBO problems through their components in the $z$-axis and the initial Hamiltonian in the $x$-axis.

The specific choice of $H_{\text{initial}} = -\Gamma \sum_i \sigma_i^x$ has the advantage that its ground state is $\ket{+}^{\otimes n}$, which can be easily prepared using Hadamard gates on the initialized qubits of $\ket{0}$. Furthermore, the inclusion of the transverse field term $H_{\text{initial}}$ in a different axis from the Hamiltonian problem is crucial, as it introduces the necessary quantum fluctuations to avoid level crossings such that the system can remain in the ground state throughout, if the time evolution is carried out slowly enough. Intuitively, the time evolution can be thought of as the very slow change of the external magnetic fields and coupling strengths in the Ising model.

\Cref{fig:AQC} shows the energy spectrum as a function of time during the adiabatic evolution, highlighting how the energy levels may change as well as showing the importance of the spectral gap between the ground state and the first excited state.

Understanding the adiabatic theorem and the role of the transverse-field Ising model is essential to appreciate how quantum-mechanical effects can be harnessed to address challenging computational problems. As it is usually computationally infeasible to calculate the spectral gap, since this would require the computation of the smallest eigenvalues for every single $t$, which itself is NP-hard, one can use machine learning techniques to optimize the time evolution speed nonetheless. This can be done exactly in the \textbf{Quantum Approximate Optimization Algorithm (QAOA)}, which discretizes the adiabatic evolution into a sequence of unitary operations that can be efficiently implemented on a quantum gate-based computer.

\section{From Adiabatic Quantum Computing to the Quantum Approximate Optimization Algorithm}
Adiabatic quantum computing (AQC) offers a powerful approach to solving combinatorial optimization problems by evolving a quantum system from the ground state of an initial Hamiltonian to the ground state of a problem Hamiltonian. This continuous time evolution relies on the adiabatic theorem, which requires the evolution to be slow enough to keep the system in its ground state.

\begin{figure}[ht]
    \centering
    \begin{subfigure}[b]{0.49\textwidth}
        \centering
        \definecolor{color_251262}{rgb}{0.890196,0.419608,0.027451}
\definecolor{color_223913}{rgb}{0.792157,0,0}
\definecolor{color_174902}{rgb}{0.592157,0,0}
\definecolor{color_29791}{rgb}{0,0,0}

\begin{tikzpicture}[scale=0.55]
    \path(0pt,0pt);

    \draw[->, thick] (0pt, -147pt) -- (265pt, -147pt) node[pos=0.5, below]{time};
    \draw[->, thick] (0pt, -147pt) -- (0pt, 20pt) node[pos=0.5, left, rotate=90, yshift=7pt, xshift=10pt] {energy};
    \draw[|-|, thin, >=stealth] (138pt, -112pt) -- (138pt, -99pt) node[left]{};
    \node at (148pt, -105pt) {$g$};

    \draw (250pt, -147pt) -- (250pt, -152pt) node[below] {$T$};
    \draw (0.5pt, -147pt) -- (0.5pt, -152pt) node[below] {$0$};

    \draw[color_251262,line width=0.6pt]
    (0pt, 2.327393pt) .. controls (22.91pt, -10.82082pt) and (46.021pt, -24.04408pt) .. (70.04pt, -30.12602pt)
     -- (70.04pt, -30.12602pt)
     -- (70.04pt, -30.12602pt)
     -- (70.04pt, -30.12602pt) .. controls (94.06pt, -36.20404pt) and (114.73pt, -40.85632pt) .. (144.31pt, -34.21581pt)
     -- (144.31pt, -34.21581pt)
     -- (144.31pt, -34.21581pt)
     -- (144.31pt, -34.21581pt) .. controls (173.9pt, -27.57529pt) and (213.42pt, -9.010132pt) .. (250.25pt, 9.638pt)
    ;

    \draw[color_223913,line width=0.6pt]
    (0pt, -15.98787pt) .. controls (10.59pt, -27.99939pt) and (39.19pt, -51.41472pt) .. (71.72pt, -59.08257pt)
     -- (71.72pt, -59.08257pt)
     -- (71.72pt, -59.08257pt)
     -- (71.72pt, -59.08257pt) .. controls (104.26pt, -66.75432pt) and (160.87pt, -71.57455pt) .. (195.74pt, -62.15283pt)
     -- (195.74pt, -62.15283pt)
     -- (195.74pt, -62.15283pt)
     -- (195.74pt, -62.15283pt) .. controls (226.52pt, -51.02408pt) and (239.03pt, -39.82867pt) .. (250.1pt, -28.9812pt)
    ;

    \draw[color_174902,line width=0.6pt]
    (0pt, -60.65547pt) .. controls (20.04pt, -74.31541pt) and (40.22pt, -88.10556pt) .. (64.52pt, -94.17577pt)
     -- (64.52pt, -94.17577pt)
     -- (64.52pt, -94.17577pt)
     -- (64.52pt, -94.17577pt) .. controls (88.82pt, -100.2538pt) and (114.94pt, -102.5819pt) .. (145.95pt, -97.24211pt)
     -- (145.95pt, -97.24211pt)
     -- (145.95pt, -97.24211pt)
     -- (145.95pt, -97.24211pt) .. controls (176.95pt, -91.90234pt) and (226.35pt, -64.36375pt) .. (250.54pt, -62.15285pt)
    ;

    \draw[color_29791,line width=0.6pt]
    (0pt, -145.2843pt) .. controls (29.21pt, -136.1321pt) and (58.43pt, -126.9916pt) .. (85.17pt, -121.4331pt)
     -- (85.17pt, -121.4331pt)
     -- (85.17pt, -121.4331pt)
     -- (85.17pt, -121.4331pt) .. controls (111.92pt, -115.8629pt) and (133.1pt, -111.7731pt) .. (160.47pt, -111.8942pt)
     -- (160.47pt, -111.8942pt)
     -- (160.47pt, -111.8942pt)
     -- (160.47pt, -111.8942pt) .. controls (187.84pt, -112.0036pt) and (235.56pt, -120.2988pt) .. (250.67pt, -122.3496pt)
    ;

\end{tikzpicture}
        \caption{Adiabatic time evolution.}
        \label{fig:AQC}
    \end{subfigure}
    \hfill
    \begin{subfigure}[b]{0.49\textwidth}
        \centering
        \definecolor{color_251262}{rgb}{0.890196,0.419608,0.027451}
\definecolor{color_223913}{rgb}{0.792157,0,0}
\definecolor{color_174902}{rgb}{0.592157,0,0}
\definecolor{color_29791}{rgb}{0,0,0}

\begin{tikzpicture}[scale=0.55]
    \path(0pt,0pt);

    \draw[->, thick] (-15pt, -147pt) -- (250pt, -147pt) node[pos=0.5, below]{time};
    \draw[->, thick] (-15pt, -147pt) -- (-15pt, 20pt) node[pos=0.5, left, rotate=90, yshift=7pt, xshift=10pt] {energy};

    \draw (235pt, -147pt) -- (235pt, -152pt) node[below] {$T$};
    \draw (-14.5pt, -147pt) -- (-14.5pt, -152pt) node[below] {$0$};

\draw[color_251262,line width=0.6pt,dash pattern=on 1.4995034pt off 2.9990067pt]
(-14.62669pt, 2.327393pt) .. controls (8.283096pt, -10.82082pt) and (31.39412pt, -24.04408pt) .. (55.41328pt, -30.12602pt)
 -- (55.41328pt, -30.12602pt)
 -- (55.41328pt, -30.12602pt)
 -- (55.41328pt, -30.12602pt) .. controls (79.43242pt, -36.20404pt) and (100.104pt, -40.85632pt) .. (129.6816pt, -34.21581pt)
 -- (129.6816pt, -34.21581pt)
 -- (129.6816pt, -34.21581pt)
 -- (129.6816pt, -34.21581pt) .. controls (159.271pt, -27.57529pt) and (198.7896pt, -9.010132pt) .. (235.621pt, 9.638pt)
;
\draw[color_223913,line width=0.6pt,dash pattern=on 1.4995034pt off 2.9990067pt]
(-14.40524pt, -15.98787pt) .. controls (-3.815567pt, -27.99939pt) and (24.58171pt, -51.41472pt) .. (57.11244pt, -59.08257pt)
 -- (57.11244pt, -59.08257pt)
 -- (57.11244pt, -59.08257pt)
 -- (57.11244pt, -59.08257pt) .. controls (89.65102pt, -66.75432pt) and (146.2633pt, -71.57455pt) .. (181.13pt, -62.15283pt)
 -- (181.13pt, -62.15283pt)
 -- (181.13pt, -62.15283pt)
 -- (181.13pt, -62.15283pt) .. controls (211.9107pt, -51.02408pt) and (224.4118pt, -39.82867pt) .. (235.482pt, -28.9812pt)
;
\draw[color_174902,line width=0.6pt,dash pattern=on 1.4995034pt off 2.9990067pt]
(-14.55566pt, -60.65547pt) .. controls (5.483078pt, -74.31541pt) and (25.66374pt, -88.10556pt) .. (49.96413pt, -94.17577pt)
 -- (49.96413pt, -94.17577pt)
 -- (49.96413pt, -94.17577pt)
 -- (49.96413pt, -94.17577pt) .. controls (74.26451pt, -100.2538pt) and (100.3813pt, -102.5819pt) .. (131.3925pt, -97.24211pt)
 -- (131.3925pt, -97.24211pt)
 -- (131.3925pt, -97.24211pt)
 -- (131.3925pt, -97.24211pt) .. controls (162.392pt, -91.90234pt) and (211.7896pt, -64.36375pt) .. (235.9806pt, -62.15285pt)
;
\draw[color_29791,line width=0.6pt,dash pattern=on 1.4995034pt off 2.9990067pt]
(-14.76527pt, -145.2843pt) .. controls (14.44518pt, -136.1321pt) and (43.66347pt, -126.9916pt) .. (70.40132pt, -121.4331pt)
 -- (70.40132pt, -121.4331pt)
 -- (70.40132pt, -121.4331pt)
 -- (70.40132pt, -121.4331pt) .. controls (97.15089pt, -115.8629pt) and (118.3302pt, -111.7731pt) .. (145.7009pt, -111.8942pt)
 -- (145.7009pt, -111.8942pt)
 -- (145.7009pt, -111.8942pt)
 -- (145.7009pt, -111.8942pt) .. controls (173.0715pt, -112.0036pt) and (220.7907pt, -120.2988pt) .. (235.8998pt, -122.3496pt)
;
\draw[color_251262,line width=0.6pt]
(-15.49576pt, 1.535034pt) -- (-15.49576pt, 1.535034pt)
 -- (-15.49576pt, 1.535034pt)
 -- (15.68343pt, 1.535034pt)
;
\draw[color_223913,line width=0.6pt]
(-15.40591pt, -16.3644pt) -- (-15.40591pt, -16.3644pt)
 -- (-15.40591pt, -16.3644pt)
 -- (15.77718pt, -16.3644pt)
;
\draw[color_174902,line width=0.6pt]
(-15.49576pt, -60.83279pt) -- (-15.49576pt, -60.83279pt)
 -- (-15.49576pt, -60.83279pt)
 -- (15.68343pt, -60.83279pt)
;
\draw[color_29791,line width=0.6pt]
(-14.7653pt, -144.7384pt) -- (-14.7653pt, -144.7384pt)
 -- (-14.7653pt, -144.7384pt)
 -- (16.41386pt, -144.7384pt)
;
\draw[color_29791,line width=0.6pt]
(16.41386pt, -135.7731pt) -- (16.41386pt, -135.7731pt)
 -- (16.41386pt, -135.7731pt)
 -- (47.59306pt, -135.7731pt)
;
\draw[color_29791,line width=0.6pt]
(47.59306pt, -126.8078pt) -- (47.59306pt, -126.8078pt)
 -- (47.59306pt, -126.8078pt)
 -- (78.77611pt, -126.8078pt)
;
\draw[color_29791,line width=0.6pt]
(78.77614pt, -119.7025pt) -- (78.77614pt, -119.7025pt)
 -- (78.77614pt, -119.7025pt)
 -- (109.9553pt, -119.7025pt)
;
\draw[color_29791,line width=0.6pt]
(109.9553pt, -114.037pt) -- (109.9553pt, -114.037pt)
 -- (109.9553pt, -114.037pt)
 -- (141.1345pt, -114.037pt)
;
\draw[color_29791,line width=0.6pt]
(141.1345pt, -111.9241pt) -- (141.1345pt, -111.9241pt)
 -- (141.1345pt, -111.9241pt)
 -- (172.3137pt, -111.9241pt)
;
\draw[color_29791,line width=0.6pt]
(171.9426pt, -113.4377pt) -- (171.9426pt, -113.4377pt)
 -- (171.9426pt, -113.4377pt)
 -- (203.1218pt, -113.4377pt)
;
\draw[color_29791,line width=0.6pt]
(203.1218pt, -117.5429pt) -- (203.1218pt, -117.5429pt)
 -- (203.1218pt, -117.5429pt)
 -- (235.9109pt, -117.5429pt)
;
\draw[color_174902,line width=0.6pt]
(16.41386pt, -80.62726pt) -- (16.41386pt, -80.62726pt)
 -- (16.41386pt, -80.62726pt)
 -- (47.59306pt, -80.62726pt)
;
\draw[color_174902,line width=0.6pt]
(47.59306pt, -94.16466pt) -- (47.59306pt, -94.16466pt)
 -- (47.59306pt, -94.16466pt)
 -- (78.77611pt, -94.16466pt)
;
\draw[color_174902,line width=0.6pt]
(78.77614pt, -99.36328pt) -- (78.77614pt, -99.36328pt)
 -- (78.77614pt, -99.36328pt)
 -- (109.9553pt, -99.36328pt)
;
\draw[color_174902,line width=0.6pt]
(109.9553pt, -100.1104pt) -- (109.9553pt, -100.1104pt)
 -- (109.9553pt, -100.1104pt)
 -- (141.1345pt, -100.1104pt)
;
\draw[color_174902,line width=0.6pt]
(141.1345pt, -94.76003pt) -- (141.1345pt, -94.76003pt)
 -- (141.1345pt, -94.76003pt)
 -- (172.3137pt, -94.76003pt)
;
\draw[color_174902,line width=0.6pt]
(171.9426pt, -84.31998pt) -- (171.9426pt, -84.31998pt)
 -- (171.9426pt, -84.31998pt)
 -- (203.1218pt, -84.31998pt)
;
\draw[color_174902,line width=0.6pt]
(201.7312pt, -72.36624pt) -- (201.7312pt, -72.36624pt)
 -- (201.7312pt, -72.36624pt)
 -- (235.9806pt, -72.36624pt)
;
\draw[color_223913,line width=0.6pt]
(13.86702pt, -39.68425pt) -- (13.86702pt, -39.68425pt)
 -- (13.86702pt, -39.68425pt)
 -- (45.05403pt, -39.68425pt)
;
\draw[color_223913,line width=0.6pt]
(45.05403pt, -55.74312pt) -- (45.05403pt, -55.74312pt)
 -- (45.05403pt, -55.74312pt)
 -- (76.2332pt, -55.74312pt)
;
\draw[color_223913,line width=0.6pt]
(76.23326pt, -63.44379pt) -- (76.23326pt, -63.44379pt)
 -- (76.23326pt, -63.44379pt)
 -- (107.4124pt, -63.44379pt)
;
\draw[color_223913,line width=0.6pt]
(107.4124pt, -66.79021pt) -- (107.4124pt, -66.79021pt)
 -- (107.4124pt, -66.79021pt)
 -- (138.5916pt, -66.79021pt)
;
\draw[color_223913,line width=0.6pt]
(139.4939pt, -67.36607pt) -- (139.4939pt, -67.36607pt)
 -- (139.4939pt, -67.36607pt)
 -- (170.6809pt, -67.36607pt)
;
\draw[color_223913,line width=0.6pt]
(170.6809pt, -64.17924pt) -- (170.6809pt, -64.17924pt)
 -- (170.6809pt, -64.17924pt)
 -- (201.864pt, -64.17924pt)
;
\draw[color_223913,line width=0.6pt]
(203.1218pt, -52.97261pt) -- (203.1218pt, -52.97261pt)
 -- (203.1218pt, -52.97261pt)
 -- (235.9109pt, -52.97261pt)
;
\draw[color_251262,line width=0.6pt]
(15.6834pt, -14.99081pt) -- (15.6834pt, -14.99081pt)
 -- (15.6834pt, -14.99081pt)
 -- (46.8665pt, -14.99081pt)
;
\draw[color_251262,line width=0.6pt]
(47.59306pt, -27.6916pt) -- (47.59306pt, -27.6916pt)
 -- (47.59306pt, -27.6916pt)
 -- (78.77611pt, -27.6916pt)
;
\draw[color_251262,line width=0.6pt]
(78.77614pt, -34.99147pt) -- (78.77614pt, -34.99147pt)
 -- (78.77614pt, -34.99147pt)
 -- (109.9553pt, -34.99147pt)
;
\draw[color_251262,line width=0.6pt]
(109.9553pt, -37.21335pt) -- (109.9553pt, -37.21335pt)
 -- (109.9553pt, -37.21335pt)
 -- (141.1345pt, -37.21335pt)
;
\draw[color_251262,line width=0.6pt]
(141.1345pt, -31.29485pt) -- (141.1345pt, -31.29485pt)
 -- (141.1345pt, -31.29485pt)
 -- (172.3137pt, -31.29485pt)
;
\draw[color_251262,line width=0.6pt]
(171.9426pt, -19.47345pt) -- (171.9426pt, -19.47345pt)
 -- (171.9426pt, -19.47345pt)
 -- (203.1218pt, -19.47345pt)
;
\draw[color_251262,line width=0.6pt]
(204.8014pt, -4.531296pt) -- (204.8014pt, -4.531296pt)
 -- (204.8014pt, -4.531296pt)
 -- (235.9806pt, -4.531296pt)
;
\end{tikzpicture}
        \caption{Approximated adiabatic time evolution.}
        \label{fig:figure2}
    \end{subfigure}
    \caption{Energy spectra for an (a) native and (b) discretized, i.e., approximated, continuous-time Hamiltonian. The differently colored lines display the different energy levels of the Hamiltonian.}
    \label{fig:adiabatictimeevolution}
\end{figure}
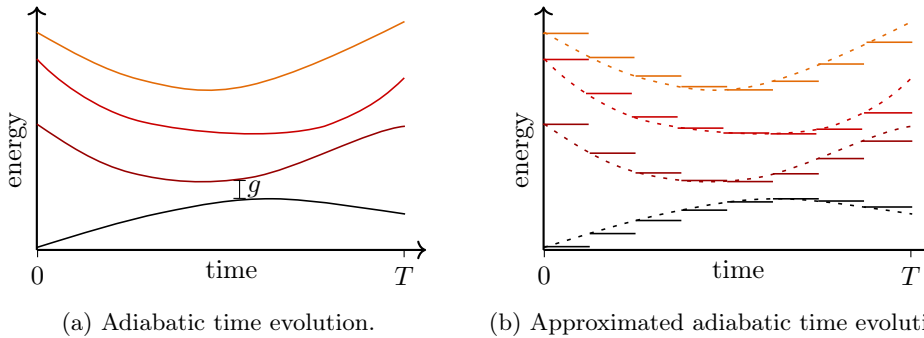

However, implementing continuous adiabatic evolution directly on a quantum-gate-based computer is practically impossible given finite time, because these computers operate using discrete unitary gates rather than evolving continuous time-dependent Hamiltonians. To simulate the adiabatic time evolution on such devices, we need to approximate the continuous evolution with a sequence of discrete operations that can be executed using quantum gates. Our goal is thus to approximate the continuous time evolution operator, which is given by the exponential of the integral of the time-dependent Hamiltonian
\begin{equation}
    U(T) = \exp\left( -i \int_0^T H(t) \, dt \right),
\end{equation}
where \( H(t) \) interpolates between the initial Hamiltonian \( H_{\text{initial}} \) and the problem Hamiltonian \( H_{\text{problem}} \) over a total time \( T \):
\begin{equation}
    H(t) = (1 - s(t)) H_{\text{initial}} + s(t) H_{\text{problem}}.
\end{equation}
Since we cannot directly implement the continuous operator \( U(T) \) on a gate-based quantum computer, we discretize the total time \( T \) into \( p \) equal intervals of length \( \delta = T / p \). Consequently, we only approximate the time evolution at discrete  points in time \( t_k = k \delta \), for \( k = 0, 1, \dots, p \). An exemplary visualizaton of the energy spectrum displaying the transition of the (discretized) time-dependent Hamiltonian is displayed in \Cref{fig:adiabatictimeevolution}.

Mathematically, this discretization is based on Gaussian quadrature (specifically, the rectangle rule) to approximate the integral
\begin{equation}
    \int_0^T H(t) \, dt \approx \sum_{k=0}^{p-1} H(t_k) \delta t.
\end{equation}
Substituting this approximation into the evolution operator, we obtain:
\begin{equation}
    U(T) \approx \exp\left( -i \sum_{k=0}^{p-1} H(t_k) \delta \right).
\end{equation}
However, since the Hamiltonians \( H(t_k) \) at different times generally do not commute (i.e., \( H(t_j) H(t_k) \ne H(t_k) H(t_j) \) for \( j \ne k \)), we cannot combine the exponentials directly. To proceed, we need a method to handle the exponential of a sum of non-commuting operators, for example the Suzuki-Trotter approximation, which provides a way to approximate the exponential of a sum of non-commuting operators by a product of exponentials of the individual operators. For two operators \( A \) and \( B \), the first-order Suzuki-Trotter formula states:
\begin{equation}
    \exp\left( A + B \right) \approx \exp(A) \exp(B) + \mathcal{O}\left( \| [A, B] \| \right),
\end{equation}
where \( [A, B] = A B - B A \) is the so-called commutator of \( A \) and \( B \), and the error depends on the magnitude of the commutator. Applying this approximation to each term in the discretized evolution operator, we have
\begin{equation}
    \exp\left( -i H(t_k) \delta \right) \approx \exp\left( -i (1 - s(t_k)) H_{\text{initial}} \delta \right) \exp\left( -i s(t_k) H_{\text{problem}} \delta \right).
\end{equation}

This approximation allows us to express the evolution operator as a product of exponentials of \( H_{\text{initial}} \) and \( H_{\text{problem}} \), which can be implemented using quantum gates corresponding to these Hamiltonians. Combining the discretization and the Suzuki-Trotter approximation, the approximate evolution operator becomes
\begin{equation}
    U(T) \approx \prod_{k=0}^{p-1} \exp\left( -i (1 - s(t_k)) H_{\text{initial}} \delta \right) \exp\left( -i s(t_k) H_{\text{problem}} \delta \right).
\end{equation}
Defining the parameters $\beta_k = (1 - s(t_k)) \delta, \quad \gamma_k = s(t_k) \delta$, we can rewrite the evolution operator as
\begin{equation}
    U(T) \approx \prod_{k=0}^{p-1} \exp\left( -i \beta_k H_{\text{initial}} \right) \exp\left( -i \gamma_k H_{\text{problem}} \right).
\end{equation}
This expression represents the total evolution as a sequence of unitary operations, each corresponding to time evolution under \( H_{\text{initial}} \) and \( H_{\text{problem}} \) for durations \( \beta_k \) and \( \gamma_k \), respectively.

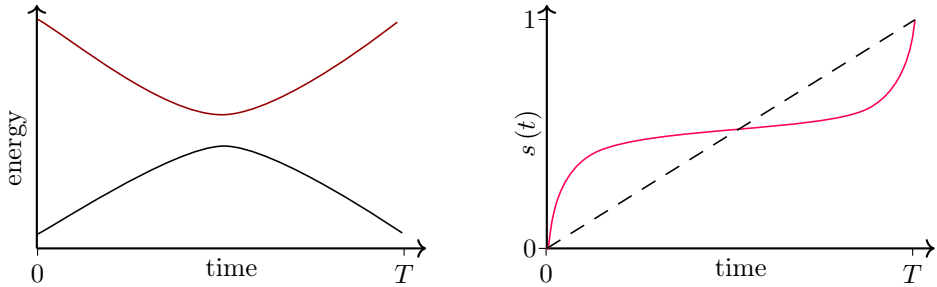
\begin{figure}[ht]
    \centering
    \begin{subfigure}[t]{0.49\textwidth}
        \centering
        \definecolor{color_174902}{rgb}{0.592157,0,0}
\definecolor{color_29791}{rgb}{0,0,0}
\begin{tikzpicture}[scale = 0.55]
\path(0pt,0pt);

\draw[->, thick] (-15pt, -147pt) -- (250pt, -147pt) node[pos=0.5, below]{time};
\draw[->, thick] (-15pt, -147pt) -- (-15pt, 20pt) node[pos=0.5, left, rotate=90, yshift=7pt, xshift=10pt] {energy};

\draw (235pt, -147pt) -- (235pt, -152pt) node[below] {$T$};
\draw (-14.5pt, -147pt) -- (-14.5pt, -152pt) node[below] {$0$};

\draw[color_174902,line width=0.6pt]
(-14.61217pt, 10.28369pt) .. controls (8.297989pt, -2.868652pt) and (69.72767pt, -54.97803pt) .. (110.548pt, -55.32568pt)
 -- (110.548pt, -55.32568pt)
 -- (110.548pt, -55.32568pt)
 -- (110.548pt, -55.32568pt) .. controls (151.3761pt, -55.67722pt) and (229.5675pt, 7.943848pt) .. (230.337pt, 8.174316pt)
;
\draw[color_29791,line width=0.6pt]
(-14.952pt, -137.2476pt) .. controls (-11.66293pt, -137.4585pt) and (77.13783pt, -76.96631pt) .. (112.216pt, -76.79834pt)
 -- (112.216pt, -76.79834pt)
 -- (112.216pt, -76.79834pt)
 -- (112.216pt, -76.79834pt) .. controls (147.3058pt, -76.62646pt) and (231.9073pt, -135.5366pt) .. (233.716pt, -136.396pt)
;
\end{tikzpicture}
        \caption{Energy spectrum for some Hamiltonian.}
        \label{fig:figure3}
    \end{subfigure}
    \hfill
    \begin{subfigure}[t]{0.49\textwidth}
        \centering
        \definecolor{color_274933}{rgb}{1,0,0.341177}
\definecolor{color_29791}{rgb}{0,0,0}
\begin{tikzpicture}[scale=.55]
\path(0pt,0pt);

\draw[->, thick] (-14pt, -147pt) -- (250pt, -147pt) node[pos=0.5, below]{time};
\draw[->, thick] (-14pt, -147pt) -- (-14pt, 20pt) node[pos=0.5, left, rotate=90, yshift=7pt, xshift=10pt] {$s\left(t\right)$};

\draw (235pt, -147pt) -- (235pt, -152pt) node[below] {$T$};
\draw (-14.5pt, -147pt) -- (-14.5pt, -152pt) node[below] {$0$};

\draw (-19.5pt, 10pt) -- (-14.5pt, 10pt) node[left] {$1$};
\draw (-19.5pt, -147pt) -- (-14.5pt, -147pt) node[left] {$0$};

\draw[color_274933,line width=0.6pt]
(-14pt, -147pt) .. controls (-10.53537pt, -152.0346pt) and (-15.89474pt, -97.26508pt) .. (20.34354pt, -80.56586pt)
 -- (20.34354pt, -80.56586pt)
 -- (20.34354pt, -80.56586pt)
 -- (20.34354pt, -80.56586pt) .. controls (56.58572pt, -63.86273pt) and (172.2459pt, -66.68304pt) .. (203.5935pt, -51.61273pt)
 -- (203.5935pt, -51.61273pt)
 -- (203.5935pt, -51.61273pt)
 -- (203.5935pt, -51.61273pt) .. controls (234.9451pt, -36.55414pt) and (235.9334pt, 8.297424pt) .. (236.6443pt, 9.836487pt)
;
\draw[color_29791,line width=0.6pt,dash pattern=on 5.9999995pt off 4.4999995pt]
(-14pt, -147pt) 
 -- (236.6443pt, 9.836441pt)
;
\end{tikzpicture}
        \caption{Optimized time evolution speed (in red) vs linear time evolution speed (dashed).}
        \label{fig:figure4}
    \end{subfigure}
    \caption{The time evolution speed in adiabatic quantum computing can be optimized based on the momentary spectral gap of the time-dependent Hamiltonian.}
    \label{fig:SpeedUpAQC}
\end{figure}

The core novelty of the Quantum Approximate Optimization Algorithm (QAOA) is treating the parameters \( \{ \beta_k, \gamma_k \} \) as adjustable variables that can be optimized to improve the algorithm's performance. This effectively allows the adaptation of the time evolution speed function $s(t)$ to the spectral properties of the Hamiltonian. More precisely, this flexibility allows us to allocate more time (larger \( \beta_k \) and \( \gamma_k \)) to regions where the spectral gap \( \Delta(t) = E_1(t) - E_0(t) \) between the ground state and the first excited state is small and less time where the gap is large. This is important, as we remember that, according to the adiabatic theorem, the evolution needs to be slow in regions where the energy gap is small to minimize transitions out of the ground state. Conversely, when the gap is large, the evolution can proceed more quickly without violating the adiabatic condition. A respective example of this is shown in \Cref{fig:SpeedUpAQC} where the time evolution speed (cf. \Cref{fig:figure4}) is varied based on the momentary spectral gap (cf. \Cref{fig:figure3}).
By treating the parameters \( \{ \beta_k, \gamma_k \} \) as independent variables rather than fixed values determined by the discretization, the Quantum Approximate Optimization Algorithm thus allows for greater flexibility in approximating the adiabatic evolution. This parameterization hence enables the algorithm to potentially achieve better performance with fewer layers (\( p \)) than would be required for a straightforward discretization.

In QAOA, the evolution is no longer strictly tied to the original adiabatic path but is instead optimized to find a good approximation to the ground state of \( H_{\text{problem}} \). This approach can be particularly advantageous when the spectral gap becomes exponentially small, as is typically the case with instances of NP-hard problems. By adjusting the parameters, we can effectively navigate the energy landscape more efficiently.

Finally, the output state of the QAOA after \( p \) layers is represented by
\begin{equation}
    |\psi_p\rangle = \prod_{k=0}^{p-1} \exp\left( -i \beta_k H_{\text{initial}} \right) \exp\left( -i \gamma_k H_{\text{problem}} \right) |\psi_0\rangle,
\end{equation}
where \( |\psi_0\rangle \) is the initial state of the system, generally chosen to be the ground state of \( H_{\text{initial}} \).

Finally, it is important to note that, throughout this derivation, we have not specified the forms of \( H_{\text{initial}} \) and \( H_{\text{problem}} \). This generality ensures that the approach is applicable to a wide range of problems and Hamiltonians used in combinatorial optimization, as we will see in the next sections.

\section{Quantum Circuit Implementation of the QAOA}
In the previous sections, we explored how the QAOA emerges from discretizing the adiabatic evolution and introducing adjustable parameters to control the evolution speed. Now, we turn our attention to the practical implementation of QAOA on a quantum gate-based computer. Specifically, we focus on constructing quantum circuits that realize the time evolution operators for the mixer and cost Hamiltonians, which are the central components of the algorithm.

To achieve this, we need to understand how to simulate Hamiltonian dynamics using quantum gates---a process known as \textbf{Hamiltonian simulation}. While Hamiltonian simulation is a broad and complex field, we will focus on the essential concepts and techniques relevant to implementing the QAOA. We also discuss how higher-order Ising interactions can be modeled with quantum gates, enabling the representation of higher-order polynomial cost functions within the algorithm.

Hamiltonian simulation involves using a quantum computer to mimic the time evolution of a quantum system governed by a Hamiltonian \( H \). The time evolution operator for a Hamiltonian \( H \) over a time \( t \) is given by the unitary operator \( U(t) = \exp( -i H t ) \). Simulating \( U(t) \) on a quantum computer allows us to study the dynamics of quantum systems and implement algorithms like the QAOA, which rely on controlled evolutions under specific Hamiltonians. In general, simulating the time evolution under an arbitrary Hamiltonian is a challenging problem due to the exponential size of the Hilbert space and the possible complexity of \( H \). However, for certain classes of Hamiltonians---particularly those that are \textbf{local}, i.e., can be expressed as sums of Hamiltonians acting on a small subset of qubits---efficient simulation methods are known. Thus, a key approach to Hamiltonian simulation involves decomposing the Hamiltonian into a sum of simpler components, often expressed in terms of \textbf{Pauli strings}. A Pauli string is a tensor product of Pauli matrices (\( I \), \( X \), \( Y \), \( Z \)) acting on different qubits. Breaking down \( H \) into these components, we can simulate the evolution under each term separately and combine the results to approximate the overall evolution. Any Hermitian operator acting on \( n \) qubits can be decomposed into a linear combination of Pauli strings (i.e., the Pauli strings form a basis of the space of Hermitian matrices):
\begin{equation}
H = \sum_k h_k P_k,
\end{equation}
where \( h_k \) are real coefficients, and \( P_k \) are Pauli strings composed of tensor products of Pauli matrices acting on the qubits. This decomposition is powerful because the exponentials of individual Pauli strings can often be implemented efficiently using quantum gates. Note, however, that decomposition of an arbitrary $n$-qubit Hamiltonian generally scales exponentially in $n$ when done classically with the standard formula of $h_k=\textnormal{tr}(P_kH)$, where $P_k\coloneqq \bigotimes_{i=1}^n \sigma_{i,k}$ and $\sigma_{i,k}\in \left\lbrace I, X, Y, Z\right\rbrace$ for all $k$.

For a Hamiltonian \( H = \sum_j H_j \), the time evolution operator $\exp\left( -i H \delta \right)$ over a short time \( \delta \) can be approximated using the \textbf{Suzuki-Trotter decomposition} via
\begin{equation}
    U(\delta) = \exp\left( -i H \delta \right) \approx \prod_j \exp\left( -i H_j \delta \right) + \mathcal{O}(\delta^2),
\end{equation}
as, while $e^{AB}\neq e^A e^B$ for non-commuting matrices $A$ and $B$, one can see that $e^{A\delta} e^{B\delta} = I + \delta(A+B) + \delta^2/2(A^2+AB+BA+B^2) + \mathcal{O}(\delta^3) \approx e^{AB \delta} = I + \delta(A+B) + \delta^2/2(A^2+2AB +B^2) + \mathcal{O}(\delta^3)$ for small $\delta\in\mathbb{R}$. This approximation becomes more accurate as \( \delta \) decreases. By applying this approach repeatedly over multiple short time steps, we can, therefore, simulate the evolution over a longer period of time.

One very important aspect of Hamiltonian simulation is that as a result of the exponentiation of the Hamiltonian, eigenvalues outside the interval $[-\pi, \pi[$ degenerate through the exponentiation as by $e^{2 \pi i (x+k)}= e^{2 \pi i x}$ for $k\in\mathbb{Z}$. This potential problem becomes clearer, if we inspect the exponentiation of a cost Hamiltonian with diagonal entries $0$ and $2\pi$. By straightforward calculation, we can see that this yields the identity operator, even though the costs for both solutions are clearly very different. Thus, when simulating Hamiltonians with eigenvalues outside the interval $[-\pi, \pi[$ (or $[0, 2\pi[$, if all eigenvalues are positive, or $[-2 \pi, 0[$, if all eigenvalues are negative), one first has to downscale the Hamiltonian with its largest eigenvalue in terms of absolute values $\max_i |\lambda_i |$. In practice, one often can put bounds on the highest/lowest costs that can occur in an optimization problem; e.g., in the Traveling Salesperson problem for $n$ cities, the largest costs would be the sum of the $n$ edges with the longest distance, and the smallest costs would be the sum of the $n$ edges with the shortest distance. In the following, we thus implicitly assume that the problem Hamiltonians are always suitably rescaled.

In the standard form of the QAOA, the mixer Hamiltonian \( H_M \) is typically chosen to be the transverse-field Ising Hamiltonian \( H_M = -\sum_{i=1}^n X_i \), where \( X_i \) is the Pauli-\( X \) operator acting on qubit \( i \). Note that we previously used the term initial Hamiltonian instead of mixer Hamiltonian for the sake of aligning with the underlying adiabatic process. The term ``mixer'' underscores that this Hamiltonian is the operator that traverses the solution space. The time evolution operator under \( H_M \) for a time \( \beta \) is \( U_M(\beta) = \exp( -i \beta H_M ) = \exp( i \beta \sum_{i=1}^n X_i ) \). Because the terms \( X_i \) acting on different qubits commute (i.e., \( X_i X_j = X_j X_i \) for \( i \neq j \)), we can factorize the exponential:
\begin{equation}
    U_M(\beta) = \prod_{i=1}^n \exp\left(i \beta X_i \right).
\end{equation}
Each exponential \( \exp( -i \beta X_i ) \) corresponds to a single-qubit rotation around the \( X \)-axis by an angle \( -2\beta \). Specifically, we use the rotation operator \( R_x(\theta) = \exp( -i \frac{\theta}{2} X ) \). So to implement \( \exp( i \beta X_i ) \), we perform a rotation \( R_x(-2\beta) \) on qubit \( i \). The quantum circuit implementation of \( U_M(\beta) \) involves applying \( R_x(-2\beta) \) to each qubit, which is straightforward and requires only single-qubit gates.

The cost Hamiltonian \( H_C \) (which we previously called problem Hamiltonian $H_{\textnormal{problem}}$ motivated from the underlying adiabatic process) encodes the objective function of the optimization problem. It is typically constructed using Ising-type interactions and can be expressed as a sum of terms involving \( Z \) operators:
\begin{equation}
    H_C = \sum_{\langle i, j\rangle} J_{ij} Z_i Z_j + \sum_{i=1}^n h_i Z_i,
\end{equation}
where \( J_{ij}\in\mathbb{R} \) are interaction strengths and \( h_i \in\mathbb{R}\) are locally applied biases. The time evolution operator under \( H_C \) for a time \( \gamma \) is defined by \( U_C(\gamma) = \exp( -i \gamma H_C ) \). Because the terms in \( H_C \) only act in the $z$-dimension for the standard Ising model, they commute. This allows the decomposition of \( U_C(\gamma) \) into a product of exponentials of the individual terms.

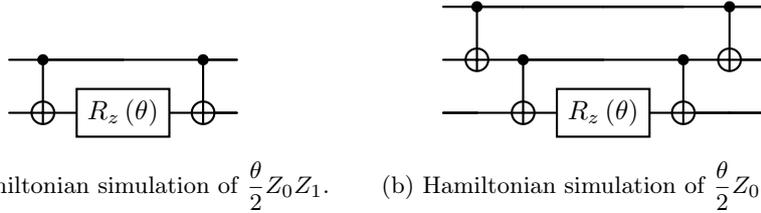
\begin{figure*}[ht!]
\centering

\begin{subfigure}[b]{0.45\textwidth}    
        \centering
        \begin{quantikz}[row sep={20pt,between origins},column sep=8pt]
\qw & \ctrl{1} & \qw & \ctrl{1} &   \qw \\
\qw & \targ{} & \gate{R_z\left(\theta\right)} & \targ{} &  \qw 
\end{quantikz}
\caption{Hamiltonian simulation of $\dfrac{\theta}{2}Z_0Z_1$.}
        \label{fig:Ham-Sim-2D}
    \end{subfigure}
    \hspace{0.2cm}
    \begin{subfigure}[b]{0.45\textwidth}
        \centering
        \begin{quantikz}[row sep={20pt,between origins},column sep=8pt]
\qw & \ctrl{1} & \qw & \qw & \qw & \ctrl{1} &\qw \\
\qw & \targ{} & \ctrl{1} & \qw & \ctrl{1} &  \targ{} &  \qw \\
\qw & \qw & \targ{} & \gate{R_z\left(\theta\right)} & \targ{} &  \qw &  \qw 
\end{quantikz}
        \caption{Hamiltonian simulation of $\dfrac{\theta}{2}Z_0Z_1Z_2$.}
\label{fig:Ham-Sim-3D}
    \end{subfigure}
    \caption{Hamiltonian simulation of the components in the cost Hamlitonian $H_C$.}
    \label{fig:Ham-Sim-Pauli-Tensors}
\end{figure*}

Consider the term \( J_{ij} Z_i Z_j \). The exponential of this term is \( U_{ij} = \exp( -i \gamma J_{ij} Z_i Z_j ) \). We can implement \( U_{ij} \) by applying a CNOT gate with qubit \( i \) as control and qubit \( j \) as target, then applying a rotation \( R_z(2\gamma J_{ij}) \) on qubit \( j \), and finally applying another CNOT gate with qubit \( i \) as control and qubit \( j \) as target. A circuit implementation of this is displayed in \Cref{fig:Ham-Sim-2D}.

For the term \( h_i Z_i \), the exponential is \( U_i = \exp( -i \gamma h_i Z_i ) \), which is a single-qubit rotation around the \( Z \)-axis by an angle \( 2\gamma h_i \), i.e., \( U_i = R_z(2\gamma h_i) \). Therefore, we apply the gate \( R_z(2\gamma h_i) \) to qubit \( i \). Combining the implementations of the two-qubit and single-qubit terms, the overall time evolution operator \( U_C(\gamma) \) can be implemented as a sequence of quantum gates. Since the terms involving different qubits of non-overlapping pairs commute, the order in which we apply these gates does not affect the result.

In some optimization problems, the cost function may involve higher-order monomials between more than two variables, e.g., $x_0 x_1 x_2$. These can be represented as higher-order Ising interactions involving products of more than two \( Z \) operators, such as \( Z_i Z_j Z_k \) for the aforementioned example of $x_0 x_1 x_2$. To implement the time evolution under such terms, we need to construct circuits that realize the exponential of these multi-qubit operators \( U_{ijk} = \exp( -i \gamma J_{ijk} Z_i Z_j Z_k ) \). One approach to implement higher-order terms is displayed in \Cref{fig:Ham-Sim-3D}. While this operation is of linear depth with respect to the number of qubits, there also exist hardware gates that can facilitate this operation in constant depth, cf. \cite{Maslov_2018}.

\begin{figure}[hbtp]
\centering
\input{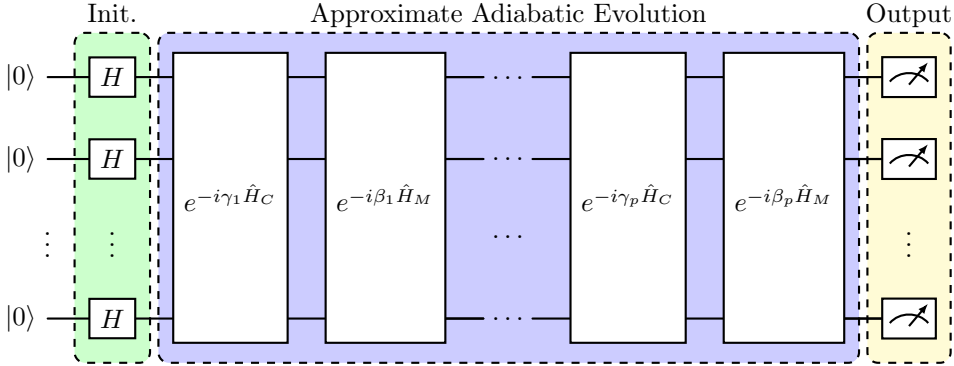}
\caption{Quantum circuit implementation of the QAOA starting in the initial state $\ket{+}^{\otimes n}$.}
\label{fig:QAOA}
\end{figure}

By combining the implementations of the mixer and cost Hamiltonians, we construct the full QAOA circuit. For each layer \( l \) of the algorithm, we apply \( U_C(\gamma_l) \) to implement the cost Hamiltonian evolution using the parameter \( \gamma_l \), and then apply \( U_M(\beta_l) \) to implement the mixer Hamiltonian evolution by applying rotations \( R_x(2\beta_l) \) to each qubit. We repeat this sequence for each of the \( p \) layers. The complete QAOA circuit consists of alternating layers of cost and mixer unitaries, parameterized by \( \{ \gamma_l, \beta_l \} \) as shown in \Cref{fig:QAOA}.

\begin{figure}[hbtp]
\centering
\input{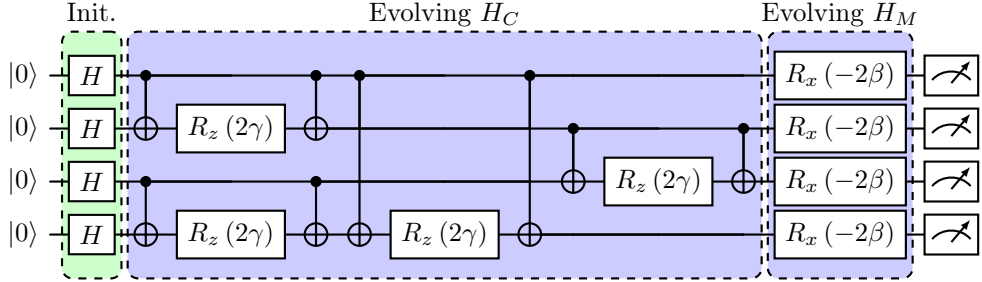}
\caption{QAOA circuit for a MaxCut problem of a four node cycle graph with $p=1$.}
\label{fig:DetailedQAOA}
\end{figure}

For a more concrete example of how the gate implementations of the cost and mixer unitary can look like in practice, we display a single QAOA layer for the typical MaxCut Hamiltonian 
\begin{equation}
    H=-\beta \sum_{i=1}^n X_i + \gamma\sum_{ij} J_{ij} Z_i Z_j
\end{equation}
of a four node cycle graph, i.e., $J_{ij}=1$ iff $j=i+1 \mod 4$ in Figure~\ref{fig:DetailedQAOA}.

\section{Training Parameters of Variational Quantum Circuits}
The QAOA relies on adjustable parameters to guide the evolution of a quantum state towards the optimal solution of a combinatorial optimization problem. These parameters, denoted as \( \{\gamma_l, \beta_l\} \) for each layer \( l \), control the time spent evolving under the cost and mixer Hamiltonians, respectively. To find the optimal set of parameters that minimize the expected value of the problem Hamiltonian, we employ gradient-based optimization methods, which require computing the derivatives of the cost function with respect to the parameters. Note that there also exist gradient-free methods to search for better parameter configurations, for which we refer to \cite{Wilson2021}.

In classical optimization, gradients are efficiently computed using techniques like backpropagation, especially in neural networks. However, in the quantum realm, computing gradients is more challenging due to the probabilistic nature of quantum measurements and the inability to directly access the quantum state. The core tool for computing these derivatives in quantum circuits is the \textbf{parameter shift rule}, which provides a way to obtain exact gradients by evaluating the circuit at the values of shifted parameters \cite{PhysRevA.98.032309}.

To understand the parameter shift rule, consider a simple example involving a single-qubit rotation around the \( Z \)-axis, represented by the gate \( R_z(\theta) = \exp(-i \theta Z / 2) \), where \( Z \) is the Pauli-\( Z \) matrix and \( \theta \) is a rotation angle. Suppose we have a quantum circuit that prepares a state \( |\psi(\theta)\rangle = R_z(\theta) |\psi_0\rangle \), where \( |\psi_0\rangle \) is some initial state, and we are interested in the expectation value of an observable \( O \) with respect to this state
\begin{equation}
\langle O \rangle (\theta) \coloneqq \langle \psi(\theta) | O | \psi(\theta) \rangle.
\end{equation}
Our goal is to compute the derivative of \( \langle O \rangle (\theta) \) with respect to \( \theta \). The parameter shift rule acknowledges that for gates generated by Hermitian operators with eigenvalues \( \pm 1 \), such as Pauli matrices, the derivative can be calculated using:
\begin{equation}
\frac{\partial \langle O \rangle (\theta)}{\partial \theta} = \frac{1}{2} \left[ \langle O \rangle \left( \theta + \frac{\pi}{2} \right) - \langle O \rangle \left( \theta - \frac{\pi}{2} \right) \right].
\end{equation}
This formula allows us to compute the exact derivative by evaluating the expectation value at two shifted angles, \( \theta + \frac{\pi}{2} \) and \( \theta - \frac{\pi}{2} \), without the need for numerical differentiation. The key insight is that the derivative of the expectation value can be expressed as a finite difference of expectation values at specific parameter shifts, which can be approximated via quantum circuit executions.

For extending this concept to general quantum circuits, consider a variational quantum circuit \( U(\boldsymbol{\theta}) \) with parameters \( \boldsymbol{\theta} = (\theta_1, \theta_2, \dots, \theta_n) \) acting on an initial state \( |\psi_0\rangle \). The circuit prepares the state \( |\psi(\boldsymbol{\theta})\rangle = U(\boldsymbol{\theta}) |\psi_0\rangle \), and we measure the expectation value of an observable \( C \). To optimize \( \langle C\rangle(\boldsymbol{\theta}) \) with respect to \( \boldsymbol{\theta} \), we need to compute the gradients \( \partial \langle C\rangle / \partial \theta_j \) for each parameter \( \theta_j \). For gates in the circuit that are parameterized by \( \theta_j \) and generated by Hermitian operators with certain spectral properties (e.g., Pauli operators), the parameter shift rule can be applied yielding
\begin{equation}
\frac{\partial \langle C\rangle}{\partial \theta_j} = r \left[ \langle C\rangle\left( \theta_j + s \right) - \langle C\rangle\left( \theta_j - s \right) \right],
\end{equation}
where \( r \) and \( s \) are constants that depend on the generator of the gate. For Pauli operators, \( r = \frac{1}{2} \) and \( s = \frac{\pi}{2} \). In the context of QAOA, the cost function is the expectation value of the problem Hamiltonian \( H_C \):
\begin{equation}
 \langle C\rangle(\boldsymbol{\gamma}, \boldsymbol{\beta}) = \langle \psi_p(\boldsymbol{\gamma}, \boldsymbol{\beta}) | H_C | \psi_p(\boldsymbol{\gamma}, \boldsymbol{\beta}) \rangle,   
\end{equation}
where \( |\psi_p(\boldsymbol{\gamma}, \boldsymbol{\beta})\rangle \) is the quantum state prepared by applying \( p \) layers of QAOA with parameters \( \boldsymbol{\gamma} = (\gamma_1, \dots, \gamma_p) \) and \( \boldsymbol{\beta} = (\beta_1, \dots, \beta_p) \). The parameters \( \gamma_l \) and \( \beta_l \) correspond to the time evolved under the cost and mixer Hamiltonians, respectively.

To compute the gradient of \( \langle C\rangle \) with respect to \( \gamma_l \), we apply the parameter shift rule:
\begin{equation}
   \frac{\partial \langle C\rangle}{\partial \gamma_l} = \frac{1}{2} \left[ \langle C\rangle\left( \gamma_l + \frac{\pi}{2}, \boldsymbol{\gamma}_{\neq l}, \boldsymbol{\beta} \right) - \langle C\rangle\left( \gamma_l - \frac{\pi}{2}, \boldsymbol{\gamma}_{\neq l}, \boldsymbol{\beta} \right) \right], 
\end{equation}
where \( \boldsymbol{\gamma}_{\neq l} \) denotes all parameters \( \gamma \) except \( \gamma_l \). An analogous expression holds for \( \partial \langle C\rangle / \partial \beta_l \).

The parameter shift rule requires us to evaluate the cost function at shifted parameter values for each parameter we wish to compute the gradient of. This means that, for each parameter, we need to perform two additional quantum circuit evaluations. Consequently, the computational cost scales linearly with the number of parameters. This is significantly more computationally intensive than classical backpropagation, where gradients with respect to all parameters can be computed simultaneously with a computational cost similar to evaluating the function itself. Despite its higher computational complexity, the parameter shift rule is currently the most efficient method known for computing exact gradients on physical quantum computers. More specifically, classical techniques like backpropagation rely on the ability to analytically compute derivatives and have direct access to the underlying computational graph, which is not possible with quantum circuits due to the no-cloning theorem and the nature of quantum measurements.

To optimize the parameters \( \boldsymbol{\gamma} \) and \( \boldsymbol{\beta} \), we use gradient-based optimization algorithms such as \textbf{gradient descent}. In gradient descent, we iteratively update the parameters in the direction opposite to the gradient of the cost function:
\begin{equation}
\theta_j \leftarrow \theta_j - \eta \frac{\partial \langle C\rangle}{\partial \theta_j},
\end{equation}
where \( \eta >0 \) is the learning rate, a hyperparameter that controls the step size of each update. By repeatedly computing the gradients and updating the parameters, we aim to find the parameter values that minimize the cost function.

It is important to note that while the optimization process is hybrid---combining classical optimization algorithms with quantum evaluations of the cost function and its gradients---the predominant computational effort lies on the quantum computer. This is because each evaluation of the cost function or its gradients requires running the quantum circuit and performing measurements to estimate expectation values. The classical optimizer processes these estimates to determine the parameter updates, but the time and resources required for this computation are typically negligible compared to the quantum circuit evaluations.

In practice, the parameter shift rule involves statistical estimation, because quantum measurements are inherently probabilistic. To obtain accurate estimates of the expectation values, a large number of measurements (shots) may be required for each circuit evaluation, further increasing the computational cost. Various techniques can be employed to mitigate this, such as using stochastic gradient descent with mini-batches of measurements or employing advanced optimization algorithms that are less sensitive to noise and estimation errors.



\section{An Example Circuit Implementation for the Maximum Cut Optimization Problem}\label{sec:implementation}
The Maximum Cut (\textit{MaxCut}) problem is a classical combinatorial optimization problem that is frequently utilized as a quantum optimization benchmark. The objective is to partition the nodes of a graph $G=(V,E)$ into two separate sets so that the number of edges between these partitions is maximal. As shown in the section on quantum annealing, we model the cost function of the minimization version of the MaxCut problem via the Hamiltonian
\begin{equation}
    H_{MC} = \sum_{(v_i,v_j)\in E}  \dfrac{1}{2}\left(\mathbb{I} - Z_i \otimes Z_j\right).
\end{equation}
Here, \( Z_i \) and \( Z_j \) are Pauli-Z operators acting on the qubits corresponding to the vertices \( i \) and \( j \) of the graph. The term \(  Z_i \otimes Z_j \) corresponds to checking whether the ends of each edge \( (i, j) \) are in different partitions by yielding a value of $+1$ when both qubits are in the same state (i.e., either \( |00\rangle \) or \( |11\rangle \)) and $-1$ when they are in opposite states (i.e., either \(|01\rangle \) or \( |10\rangle \)), effectively counting the edge as cut. This cost function corresponds directly to the number of cuts, but shifted by a constant.

We now show how to implement the QAOA in the \texttt{Python} programming language using the \texttt{pennylane} quantum circuit simulator.

\begin{lstlisting}[title=Imports]
import pennylane as qml
from pennylane import numpy as np
import networkx as nx
from matplotlib import pyplot as plt
import random
import numpy
\end{lstlisting}

\begin{figure}[H]
    \includegraphics[width=0.6\columnwidth]{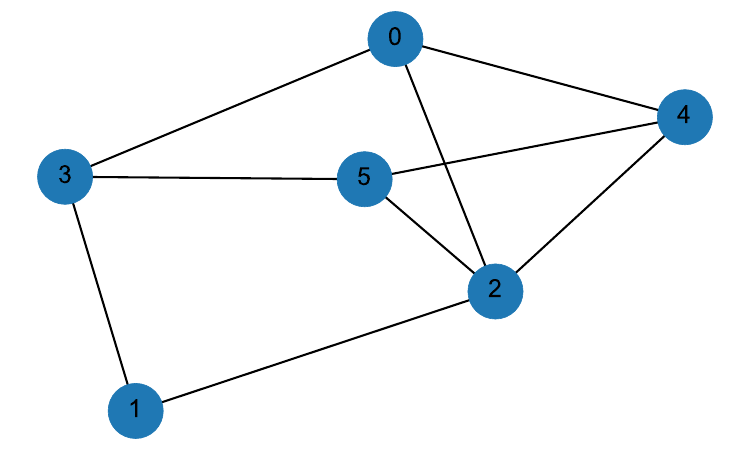}
\centering
\caption{Randomly generated Erdos-Rényi graph sample.}
\label{fig:graph_example}
\end{figure}

With the imports in place, we can use the \texttt{networkx} graph library to construct a graph sample to be used for this example run. As a simple example, we create an \textit{Erdos-Rényi} graph with $n=6$ vertices that are interconnected randomly with the specified probability $p=0.5$. In addition to that, we specify global seeds to ensure reproducibility. The resulting graph is shown in \Cref{fig:graph_example}.

\begin{lstlisting}[title=Example Graph]
random.seed(42)
numpy.random.seed(42)
n = 6
G = nx.fast_gnp_random_graph(n, 0.5)
plt.figure(figsize=(5, 3))
pos = nx.spring_layout(G)
nx.draw(G, with_labels=True, node_size=700, pos=pos)
plt.show()
\end{lstlisting}

Next, we assemble the QAOA circuit by defining the mixer and cost layers through the built-in \texttt{pennylane} function for the MaxCut problem. However, in \texttt{pennylane} as well as in other frameworks, it is assumed that the negation in the mixer Hamiltonian $H_M=\sum_i - X_i$ is absorbed into $\beta$, which we also do here. Further, we also have to introduce the rescaling factor for the Hamiltonian simulation to avoid degeneracies in the eigenvalues of the cost operator. For the MaxCut problem, this can be done by dividing with the maximum possible cut size, which can be bounded by the total number of edges in the graph.

\begin{lstlisting}[title=Completed Circuit]
depth = 10
wires = range(n)

H_C, H_M = qml.qaoa.maxcut(G)
def U_M(beta):
    qml.qaoa.mixer_layer(beta, H_M/G.number_of_edges())
def U_C(gamma):
    qml.qaoa.cost_layer(gamma, H_C/G.number_of_edges())

# define a single qaoa layer
def qaoa_layer(gamma, beta):
    U_C(gamma)
    U_M(beta)

# define the qaoa circuit
def qaoa_circuit(params):
    # Initialize ground state of mixer Hamiltonian
    for wire in wires:
        qml.Hadamard(wires=wire)
    # Apply qaoa layers
    qml.layer(qaoa_layer, depth, params[0], params[1])
\end{lstlisting}

\begin{figure}[ht]
    \includegraphics[width=.9\columnwidth]{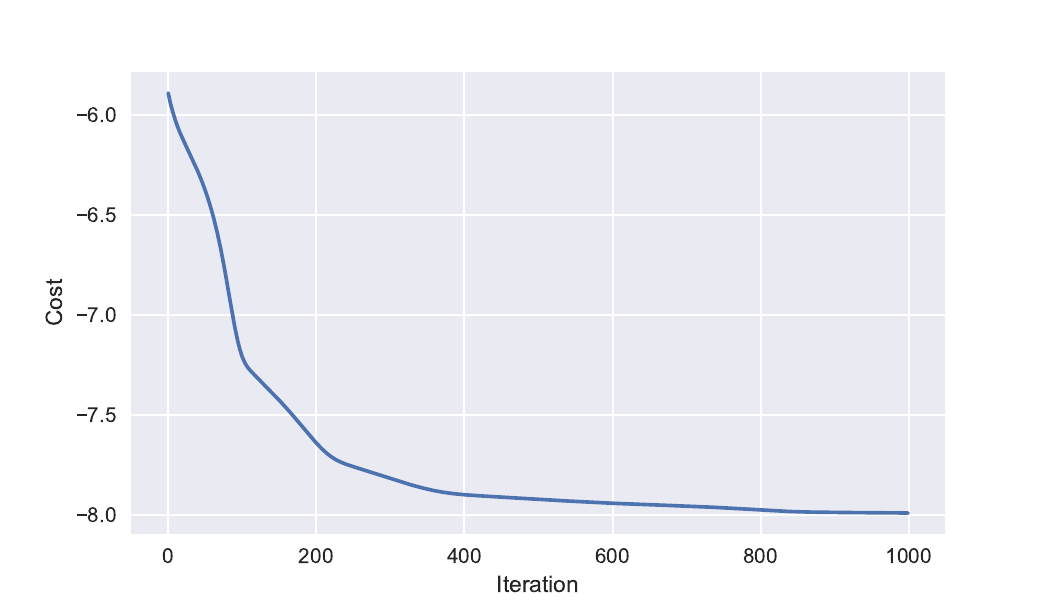}
\centering
\caption{Optimization process of random MaxCut graph instance shown in Fig~\ref{fig:graph_example}.}
\label{fig:graph_example_plot}
\end{figure}

The MaxCut objective that we have previously defined is translated into code, and the resulting optimization using a specialized gradient descent optimizer is shown in \Cref{fig:graph_example_plot}. 

\begin{lstlisting}[title=MaxCut Cost Function]
dev = qml.device("default.qubit", wires=wires)

@qml.qnode(dev)
def cost_fun(params):
    qaoa_circuit(params)
    return qml.expval(H_C)
\end{lstlisting}

\begin{lstlisting}[title=Whole Optimization Loop]
optimizer = qml.AdamOptimizer()
steps = 1000
T = 7.5
beta = [- (1 - i/depth)* T/depth for i in range(depth)]
gamma = [(i/depth)*T/depth for i in range(depth)]
params = np.array([beta, gamma], requires_grad=True)
cost_list = []
initial = True

# start optimization
for step in range(steps):
    params, cost = optimizer.step_and_cost(cost_fun, params)
    cost_list.append(cost)
    if initial:
        print(f"Initial cost: {cost}")
        initial = False
    print(f"Cost at step {step}: {cost}", end="\r")
plt.figure(figsize=(7, 4))
plt.plot(cost_list)
plt.ylabel("Cost")
plt.xlabel("Iteration")
plt.show()
\end{lstlisting}

\begin{figure}[ht]
    \includegraphics[width=.9\columnwidth]{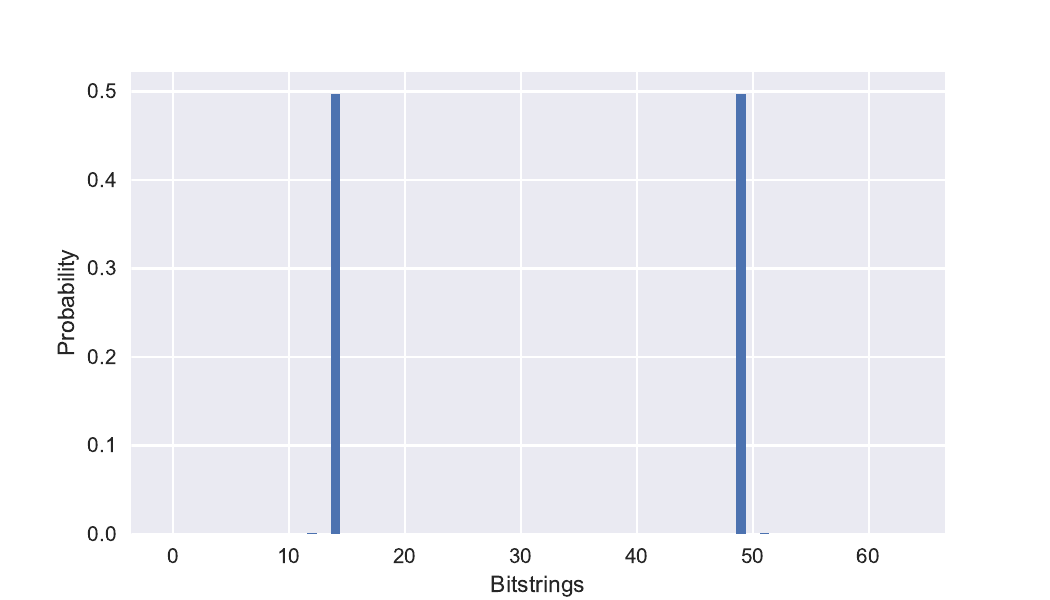}
\centering
\caption{Measured outcomes after parameter optimization.}
\label{fig:probs}
\end{figure}

Having computed suitable values for the parameters $\beta$ and $\gamma$, we can clearly see the convergence towards two (optimal) solutions when measuring the results of a QAOA circuit execution with these parameter values in \Cref{fig:probs}.

\begin{lstlisting}[title=Probability Function]
@qml.qnode(dev)
def probability_circuit(gamma, beta):
    qaoa_circuit([gamma, beta])
    return qml.probs(wires=wires)

probs = probability_circuit(params[0], params[1])
plt.figure(figsize=(7, 4))
plt.style.use('seaborn-v0_8')
plt.bar(range(2 ** len(wires)), probs)
plt.ylabel("Probability")
plt.xlabel("Bitstrings")
plt.show()
\end{lstlisting}

\begin{figure}
     \centering
     \begin{subfigure}[b]{0.48\textwidth}
         \centering
         \includegraphics[width=\textwidth]{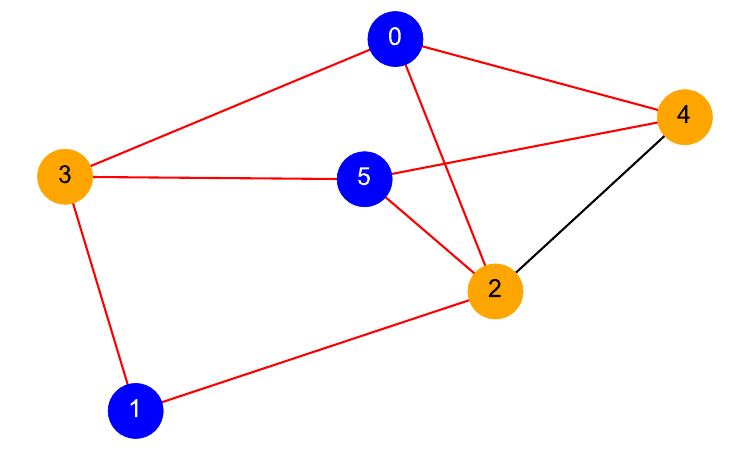}
         \caption{Cut for the bitstring $001110$.}
         \label{fig:graph_example_solution1}
     \end{subfigure}
     \hfill
     \begin{subfigure}[b]{0.48\textwidth}
         \centering
         \includegraphics[width=\textwidth]{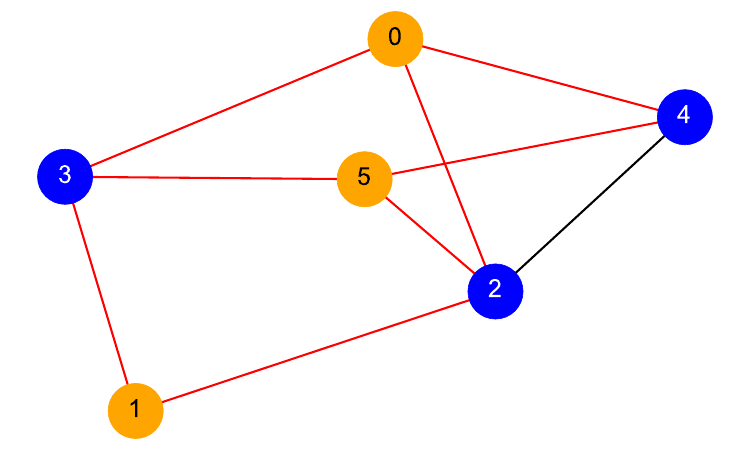}
         \caption{Cut for the bitstring $110001$.}
         \label{fig:graph_example_solution2}
     \end{subfigure}
    \caption{Resulting optimal binary solutions visualized with cut edges in red.}
    \label{fig:graph_example_solutions}
\end{figure}

Finally, we can translate these bitstrings back to partitions of the graph showing the two results found with the highest probability, which also correspond to the global optima for this problem. These results are displayed in \Cref{fig:graph_example_solutions}.

\section{Incorporating Constraints into the QAOA}
In many combinatorial optimization problems, we encounter constraints that must be satisfied for a solution to be considered feasible. Traditional approaches in the QAOA incorporate these constraints into the cost function using penalty terms. However, this method can lead to complex cost Hamiltonians and may require careful tuning of penalty weights to ensure that feasible solutions are favored. Even then, empiric experiments frequently display seemingly impenetrable difficulties for highly constrained problems with this method. An alternative approach involves the use of \textbf{Grover mixers}, which allow the integration of specific constraints by restricting the evolution of the quantum state to the subspace of feasible solutions.

The key idea behind Grover mixers is to design the initial state and the mixer Hamiltonian such that the quantum evolution remains entirely within the subspace of valid states, i.e., those that satisfy all the constraints. This ensures that infeasible states are never populated during the execution of the algorithm. To achieve this, we start by preparing an initial state that is an equal superposition over all valid states. The Grover mixer is then constructed to only map valid states onto other valid states, preserving the feasibility of the solutions throughout the algorithm.

Formally, let \( \mathcal{H}_{\text{valid}} \) denote the subspace of the Hilbert space spanned by the valid computational basis states \( \{ |x\rangle \} \) that satisfy the constraints of the problem. The initial state \( |F\rangle \) is prepared as
\begin{equation}
    |F\rangle = \frac{1}{\sqrt{|F|}} \sum_{x \in F} |x\rangle,
\end{equation}
where \( F \subseteq \{0,1\}^n \) is the set of valid bit strings, and \( |F| \) is the number of valid states. This state is an equal superposition over all valid solutions. The cost Hamiltonian \( H_C \) is designed to encode the objective function, assigning lower energies to states corresponding to better solutions within \( \mathcal{H}_{\text{valid}} \).

The Grover mixer \( H_M \) is constructed to only induce transitions between valid states. It acts as a quantum equivalent of a Markov chain that preserves the set \( F \). Formally, the Grover mixer is defined as:
\begin{equation}
H_M = \sum_{x,y \in F} M_{xy} |x\rangle \langle y|,
\end{equation}
where \( M_{xy} \in \mathbb{R}\) can be regarded as matrix elements defining the transitions between valid states, which can vary depending on the desired mixing operation. Importantly, the specific form of \( H_M \) ensures that \( H_M \) only has support within \( \mathcal{H}_{\text{valid}} \) and that it is Hermitian. Furthermore, for the standard choice of $M_{xy}=1$, it is easy to verify that $\ket{F}$ is the ground state of $-H_M$, analogously to how $\ket{+}^{\otimes n}$ is the ground state of $-\sum_i X_i$. Thus, for a circuit implementation of the Grover mixer QAOA, the negative sign also has to be absorbed into $\beta$, just as in the \texttt{pennylane} implementation of the MaxCut problem in \Cref{sec:implementation}.

\begin{figure}[hbtp]
\centering
\input{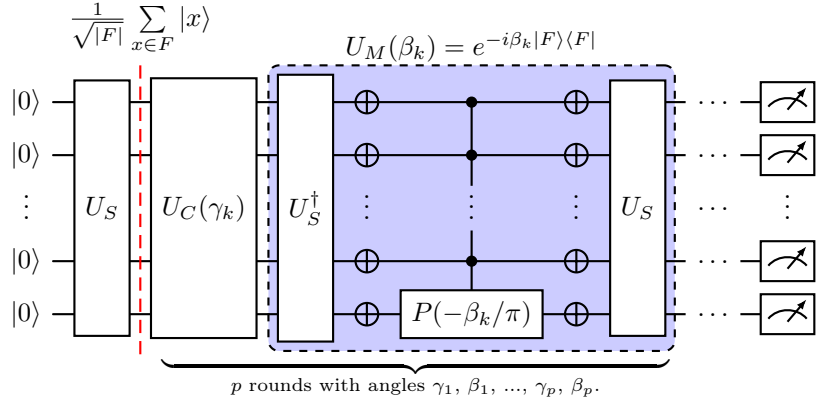}
\caption{Grover mixer QAOA. The $P(\varphi)$ gate denotes a rotation around the $z$-axis in the Bloch sphere with a shifted global phase compared to the $R_z$ gate, i.e., $P(\varphi):\ket{0}\mapsto\ket{0}$ and $P(\varphi):\ket{1}\mapsto e^{i\varphi}\ket{1}$.}
\label{fig:groverQAOA}
\end{figure}

In practice, the unitary corresponding to the time evolution of the Grover mixer can be built using reflections over the initial state and the uniform superposition of valid states, i.e.,
\begin{align}
    e^{-i \beta H_M} &=\sum_{k=0}^{\infty} \frac{(-i \beta)^k (|F\rangle \langle F|)^k}{k!}=I + \sum_{k=1}^{\infty} \frac{(-i \beta)^k}{k!} |F\rangle ((\langle F|F\rangle)^{k-1} \langle F|)\\
    &= \mathbb{I} - \left(1 - e^{-i\beta}\right) |F\rangle \langle F| =U_S \left(\mathbb{I} - \left(1 - e^{-i\beta}\right) |0\rangle \langle 0|\right) U_S^\dagger,
\end{align}
if $M_{xy}=1$ for all $x,y\in F$. This allows for a straightforward quantum circuit implementation of the Grover mixer when having access to the unitary operator implementing the state preparation $U_S$, as displayed in \Cref{fig:groverQAOA}. The name ``Grover'' mixer becomes clear when setting $\beta = \pi$, as the mixer unitary takes the form of a Grover operator used in Grover's algorithm, i.e., $e^{-i \beta H_M}=\mathbb{I}-2|F\rangle \langle F|$. For readers familiar with Grover's algorithm, it becomes apparent that the alternating applications of the cost unitary and the Grover mixer is analogous to amplitude amplification, where the amplitudes of desirable states are increased through subsequent reflections. Another very useful consequence of the Grover mixer is that, if the initial state is an equal superposition over all valid states, then each eigenstate with equal costs will also have equal measurement probability under an adiabatic time evolution.

To illustrate how Grover mixers can be constructed for specific constraints, consider the case of a \textbf{one-hot encoding}, where exactly one qubit in a group of \( n \) qubits is in the state \( |1\rangle \), and the rest are in the state \( |0\rangle \). The valid states in this encoding are the \textbf{Dicke states} with Hamming weight one. The initial state is prepared as:
\begin{equation}
    |F\rangle = \frac{1}{\sqrt{n}} \sum_{i=1}^n |e_i\rangle,
\end{equation}
where \( |e_i\rangle \) denotes the computational basis state with a \( 1 \) in the \( i \)th position and \( 0 \) elsewhere. The Grover mixer for one-hot encoding is designed to perform rotations within the subspace spanned by the Dicke states. While the explicit construction of these circuits can be complex, algorithms for preparing Dicke states with circuit depth linear in the number of qubits exist and can be utilized. For details on Dicke states, see Ref.~\cite{Dicke}.

In summary, Grover mixers provide a powerful method for incorporating hard constraints into QAOA without relying on penalty terms in the cost function. By initializing the quantum state in an equal superposition of valid states and using a mixer Hamiltonian that only allows transitions within the feasible subspace, we can efficiently explore the space of feasible solutions and enhance the probability of finding optimal solutions that satisfy all constraints. For more details on Grover mixers, see Ref.~\cite{GroverMixer}.

\section{Applying QAOA to Real-World Optimization Problems: A Step-by-Step Guide}

As outlined in the previous sections, implementing the QAOA for practical use cases involves a sequence of carefully orchestrated steps. Each phase requires attention to detail and an understanding of both quantum and classical computational principles. In the following, we summarize the essential stages of applying QAOA to a real-world optimization problem, providing a comprehensive guide for practitioners.

\subsection{Formulating the Optimization Problem as a Binary Polynomial}
The first step in applying the QAOA is to translate the given optimization problem into a mathematical form suitable for quantum computation. This typically involves expressing the problem as a binary polynomial, where variables are represented by binary bits \( x_i \in \{0, 1\} \) or spin variables \( s_i \in \{-1, +1\} \). The objective function, along with any constraints, must be encoded into a cost function that the quantum algorithm can process.

For unconstrained problems, the objective function can be directly mapped onto a cost Hamiltonian \( H_C \) by expressing it as a sum of Pauli-\( Z \) operators and their products. For instance, in the case of the MaxCut problem, the cost Hamiltonian can be written as:

\begin{equation}
H_C = \sum_{\langle i,j \rangle} \frac{1}{2}(\mathbb{I} - Z_i Z_j),
\end{equation}

where \( \langle i,j \rangle \) denotes the edges of a graph, and \( Z_i \) and \( Z_j \) are Pauli-\( Z \) operators acting on qubits representing vertices \( i \) and \( j \).

When constraints are present, they can be incorporated either by adding penalty terms to the cost function or by employing methods like Grover mixers. Penalty terms increase the energy of states that violate the constraints, discouraging the algorithm from exploring infeasible solutions. However, this can complicate the cost Hamiltonian and may require tuning penalty weights. Alternatively, Grover mixers can allow for the efficient integration of constraints by restricting the quantum evolution to the subspace of feasible solutions. This approach involves preparing an initial state that is an equal superposition of all valid states and designing a mixer Hamiltonian that only transitions between these states, ensuring that infeasible solutions are never populated.

\subsection{Constructing the Quantum Circuits for the Cost and Mixer Unitaries}
With the problem formulated, the next step is to construct the quantum circuits that implement the cost and mixer unitaries. The cost unitary \( U_C(\gamma) = \exp(-i \gamma H_C) \) requires simulating the time evolution under the cost Hamiltonian. This involves decomposing \( H_C \) into sums of Pauli strings and using Hamiltonian simulation techniques to implement the corresponding exponentials. For this, one can simply formulate the cost function as a binary polynomial and then translate it to an Ising Hamiltonian via mapping $x_i \in \lbrace 0,1\rbrace$ onto $(I-Z_i)/2$ or alternatively $s_i \in \lbrace -1,1\rbrace$ onto $Z_i$.

The mixer unitary \( U_M(\beta) \) is designed based on the choice of mixer Hamiltonian \( H_M \). For standard QAOA, the mixer is often a transverse-field Hamiltonian \( H_M = \sum_i X_i \), leading to simple single-qubit rotations in the circuit. However, when using Grover mixers to enforce constraints, the mixer may involve more complex operations. For example, in the case of a one-hot encoding, where the valid states are Dicke states with a fixed Hamming weight, the mixer unitary must be constructed to perform rotations within the subspace of these states. This can be achieved using Dicke state preparation techniques and implementing unitaries that permute amplitudes among the valid states without introducing infeasible ones.

\subsection{Specifying the Algorithm Parameters}
Before running the algorithm, we need to specify the parameters that govern its evolution. This includes choosing the number of QAOA layers \( p \), which determines the depth of the circuit and the algorithm's expressive power. A larger \( p \) allows for a more accurate approximation of the adiabatic evolution but requires more computational resources and is more susceptible to noise in noisy intermediate-scale quantum (NISQ) hardware.

Initial values for the parameters \( \boldsymbol{\beta} = (\beta_1, \beta_2, \dots, \beta_p) \) and \( \boldsymbol{\gamma} = (\gamma_1, \gamma_2, \dots, \gamma_p) \) must also be selected. These parameters control the duration of evolution under the mixer and cost Hamiltonians at each layer. While random initialization is possible, informed choices based on problem-specific insights or heuristic methods can lead to faster convergence during the optimization process.

\subsection{Mapping the Quantum Circuit onto Physical Hardware}
An essential aspect of implementing the QAOA is ensuring that the logical quantum circuit can be executed on the target quantum hardware. Quantum devices have specific architectures with limitations such as qubit connectivity, coherence times, and native gate sets. Qubits in the hardware may not be fully connected, meaning that certain two-qubit gates can only be applied between physically adjacent qubits. To address this, we may need to insert SWAP gates into the circuit to bring qubits into proximity for interactions, effectively reordering qubits to satisfy connectivity constraints.

Additionally, the logical gates used in the circuit must be decomposed into sequences of hardware-native gates. For example, a controlled-phase gate might be implemented using a combination of CNOT and single-qubit rotation gates. This process, known as gate decomposition or transpilation, optimizes the circuit for the specific capabilities and limitations of the hardware.

Software frameworks like \texttt{qiskit} and \texttt{pennylane} provide functionalities to automate much of this mapping process. They include features for circuit optimization, qubit routing, and error mitigation techniques. These frameworks can analyze the logical circuit, adapt it to the hardware topology, and optimize it to reduce depth and gate count, which is crucial to maintaining fidelity in NISQ devices.

\subsection{Training the Parameters Using Optimization Algorithms}
With the hardware-specific circuit prepared, the next phase involves training the parameters \( \boldsymbol{\beta} \) and \( \boldsymbol{\gamma} \) to minimize the expected value of the cost Hamiltonian. This is achieved through iterative optimization algorithms that adjust the parameters based on the measured outputs from the quantum circuit. An important ingredient in this context is the choice of initial parameter values (e.g., according to the discretized adiabatic process underlying the QAOA).

Gradient-based optimization can be considered the standard method for finding suitable parameter values. Here, the gradient of the cost function with respect to the parameters is used to guide the updates. Techniques like gradient descent or the Adam optimizer (which uses an adaptive learning rate) can be applied, leveraging the parameter shift rule to compute the necessary gradients. The parameter shift rule allows for the exact calculation of gradients by evaluating the quantum circuit at shifted parameter values. Although this method can be computationally intensive due to the need for multiple circuit evaluations, it provides accurate gradient estimates that are essential for effective optimization.

Alternatively, gradient-free optimization methods like Nelder-Mead or Bayesian optimization can be used, especially when gradient calculation is impractical or when dealing with noisy measurements. These methods rely on evaluating the cost function at various parameter points and updating the parameters based on heuristic rules or probabilistic models.

\subsection{Executing the Algorithm and Benchmarking Results}
After the parameters have been optimized, the final step is to run the QAOA circuit with the trained parameters and collect the measurement results. The results provide a solution or a set of candidate solutions to the optimization problem. Due to the probabilistic nature of quantum measurements, the algorithm may need to be run multiple times to obtain statistically significant results.

Evaluating the performance of QAOA involves benchmarking the obtained solutions against those produced by classical algorithms. Exact methods like branch-and-bound or branch-and-cut solvers (e.g., CPLEX, Gurobi) provide optimal solutions for comparison, albeit often with higher computational costs for large problems. Metaheuristic algorithms such as simulated annealing or genetic algorithms offer alternative benchmarks, representing the current practical approaches for tackling complex optimization tasks.

By comparing metrics such as solution quality, computational time, and resource consumption, we can assess the advantages and limitations of QAOA in the specific context of the problem. This analysis is crucial for understanding the potential of quantum algorithms in solving real-world optimization problems and for identifying areas where further improvements or adaptations are necessary. Interesting outcomes of such benchmarks are, for example, that the Traveling Salesperson Problem is best solved using suitable Grover mixers with the QAOA~\cite{bucher2024robustbenchmarkingquantumoptimization} and that problems with higher-order polynomial cost functions are likely to be solved better using the QAOA than their quadratized form necessary for D-Wave's Quantum Annealing~\cite{10.1145/3583133.3596358}.

\section{The Variational Quantum Eigensolver as a Generalization of the QAOA}
The Quantum Approximate Optimization Algorithm (QAOA) can be viewed as a search algorithm navigating the exponentially large Hilbert space of a quantum system to find the ground state corresponding to the optimal solution of a combinatorial optimization problem. By discretizing the adiabatic evolution and employing parameterized quantum circuits, QAOA explores specific pathways within this vast space. Building upon this concept, the \textbf{Variational Quantum Eigensolver (VQE)} offers a more general framework for approximating the ground state of a given Hamiltonian, providing flexibility in the choice of quantum circuits and optimization strategies. More concretely, the VQE leverages the variational principle, which states that for any parameter dependent quantum state $|\psi(\boldsymbol{\theta})\rangle$, the expectation value of the Hamiltonian provides an upper bound to the ground state energy $E_0$:
\begin{equation}
E(\boldsymbol{\theta}) = \langle \psi(\boldsymbol{\theta}) | H | \psi(\boldsymbol{\theta}) \rangle \geq E_0.    
\end{equation}
Just like the QAOA, the VQE works by iteratively adjusting the parameters $\boldsymbol{\theta}$ to minimize $E(\boldsymbol{\theta})$, thereby approximating the ground state energy and the corresponding eigenstate.

The core idea of VQE is to use a parameterized quantum circuit (PQC) to explore the Hilbert space and approximate the ground state of $\hat{H}$. By tuning the parameters $\boldsymbol{\theta}$, the PQC generates a family of trial states $|\psi(\boldsymbol{\theta})\rangle = U(\boldsymbol{\theta}) |\psi_0\rangle$ that sample different regions of the Hilbert space. While VQE was initially proposed for quantum chemistry applications, it can be generalized to solve arbitrary combinatorial optimization problems by following these steps:
\begin{enumerate}
    \item \textbf{Problem Encoding}: Map the combinatorial optimization problem defined by a cost function $f: X \rightarrow \mathbb{R}$ onto a Hamiltonian $\hat{H}$. This involves defining an encoding $e: X \rightarrow \{0,1\}^n$ that translates problem variables into binary representations. The Hamiltonian is constructed such that its diagonal elements correspond to the cost function evaluated at the encoded states $
    H = \sum_{x \in X} f(x) | e(x) \rangle \langle e(x) |$. Finding the ground state of $G$ is equivalent to finding the global minimum of $f$.
    \item \textbf{Circuit Ansatz Selection}: Choose a parameterized quantum circuit $U(\boldsymbol{\theta})$ that serves as a function approximator mapping the initial state to the ground state. The ansatz should therefore be able to map the given initial state onto a subspace of the Hilbert space that contains states with close-to-ground-state energies given suitable parameters.
    \item \textbf{Initial State Preparation}: Initialize the quantum system in a suitable state $|\psi_0\rangle$. This is typically the all-zero state $|0\rangle^{\otimes n}$ for simplicity. The initial parameters $\boldsymbol{\theta}$ can be set to zero or randomized.
    \item \textbf{Optimization Strategy}: Select an appropriate classical optimization algorithm like Adam to adjust the parameters $\boldsymbol{\theta}$ based on measurements from the quantum circuit. The optimizer aims to minimize the expectation value $E(\boldsymbol{\theta})$.
\end{enumerate}

Ideally, one might consider using the most general circuit possible, capable of implementing any unitary transformation for an $n$-qubit system, i.e., any $U$ from the special unitary group $SU(2^n)$. Such a circuit would, in principle, allow access to any state in the Hilbert space. However, practical implementations face significant challenges:
\begin{itemize}
    \item \textbf{Barren Plateaus}: Highly expressive circuits tend to produce regions in the parameter space where the gradient of the cost function becomes exponentially small---known as barren plateaus. This phenomenon hampers the trainability of the circuit because optimization algorithms rely on gradient information to update parameters effectively.
    \item \textbf{Expressibility and the Haar Measure}: The expressibility of a PQC refers to its ability to uniformly explore the unitary space, akin to sampling from the Haar measure---the unique uniform measure over the unitary group. While high expressibility might seem desirable, it correlates with an increased likelihood of encountering barren plateaus. Circuits that closely approximate Haar-random unitaries distribute probability amplitudes too evenly, making it difficult for optimization algorithms to find directions that significantly reduce the cost function.
\end{itemize}
Given these challenges, it becomes crucial to design ansätze that balance expressibility and trainability. The ansatz thus should be tailored to the specific given problem to capture relevant features without introducing unnecessary complexity. Finding a suitable ansatz for a given problem is a central and ongoing area of research in quantum computing. A well-designed ansatz can mitigate issues related to barren plateaus and improve the efficiency of the VQE algorithm. Several approaches have been proposed:
\begin{itemize}
    \item \textbf{Problem-Inspired Ansätze}: These circuits are constructed based on the structure of the Hamiltonian or the nature of the problem. For example, the Unitary Coupled Cluster ansatz in quantum chemistry incorporates domain knowledge about electron interactions.
    \item \textbf{Hardware-Efficient Ansätze}: These circuits are designed to match the connectivity and gate set of the quantum hardware, minimizing circuit depth and gate errors. While they are easy to implement, they may suffer from expressibility-trainability trade-offs.
    \item \textbf{Layered Ansätze with Entangling Gates}: By strategically placing entangling gates and adjusting circuit depth, one can create ansätze that are both expressive and trainable for certain problems.
\end{itemize}
Despite these developments, there is no one-size-fits-all approach for ansatz design. The effectiveness of an ansatz often depends on the specific problem, the properties of the Hamiltonian, and the capabilities of the quantum hardware. A promising idea in this context is using classical machine learning techniques to discover and optimize quantum circuit ansätze:
\begin{itemize}
    \item \textbf{Evolutionary Algorithms}: By treating circuit structures as individuals in a population, evolutionary strategies can iteratively improve ansätze through selection, mutation, and crossover operations.
    \item \textbf{Reinforcement Learning}: Agents are trained to construct circuits by maximizing a reward function related to the performance of the VQE algorithm.
    \item \textbf{Neural Architecture Search}: Deep learning methods are employed to search over the space of possible circuit architectures, guided by performance metrics.
\end{itemize}
These approaches aim to automate the design of ansätze, potentially uncovering novel circuit structures that are well-suited to specific problems. While initial results are promising, challenges remain in scaling these methods and ensuring that the discovered circuits are practical for implementation on quantum hardware.

In summary, the VQE extends the principles of QAOA by providing a flexible framework for approximating ground states of arbitrary Hamiltonians. By leveraging parameterized quantum circuits and hybrid optimization techniques, VQE opens avenues for solving a wide range of problems in quantum chemistry, materials science, and combinatorial optimization. However, the success of VQE critically depends on the choice of ansatz. Balancing expressibility and trainability is essential to avoid issues like barren plateaus and to ensure efficient convergence. As quantum hardware continues to advance, and as research into ansatz design progresses, VQE holds the potential to become a powerful tool in the quantum computing repertoire.

\section*{Acknowledgements}
The project underlying this report was funded by the German Federal Ministry of Research, Technology and Space under the funding code 01MQ22008A. The sole responsibility for the manuscript’s contents lies with the authors. 

\bibliography{sn-bibliography}

@article{lucas2014ising,
author = {Lucas, Andrew},
year = {2014},
month = {02},
pages = {5},
title = {Ising formulations of many NP problems},
volume = {2},
journal = {Frontiers in Physics},
doi = {10.3389/fphy.2014.00005}
}

@article{Maslov_2018,
doi = {10.1088/1367-2630/aaa398},
year = {2018},
month = {mar},
publisher = {IOP Publishing},
volume = {20},
number = {3},
pages = {033018},
author = {Dmitri Maslov and Yunseong Nam},
title = {Use of global interactions in efficient quantum circuit constructions},
journal = {New Journal of Physics}
}

@InProceedings{Dicke,
author="B{\"a}rtschi, Andreas
and Eidenbenz, Stephan",
editor="G{\k{a}}sieniec, Leszek Antoni
and Jansson, Jesper
and Levcopoulos, Christos",
title="Deterministic Preparation of Dicke States",
booktitle="Fundamentals of Computation Theory",
year="2019",
publisher="Springer International Publishing",
address="Cham",
pages="126--139",
isbn="978-3-030-25027-0",
doi="10.1007/978-3-030-25027-0_9"
}

@INPROCEEDINGS{GroverMixer,
  author={Bärtschi, Andreas and Eidenbenz, Stephan},
  booktitle={2020 IEEE International Conference on Quantum Computing and Engineering (QCE)}, 
  title={Grover Mixers for QAOA: Shifting Complexity from Mixer Design to State Preparation}, 
  year={2020},
  volume={},
  number={},
  pages={72-82},
  doi={10.1109/QCE49297.2020.00020}}

@article{PRXQuantum.5.030348,
  title = {Solving Boolean Satisfiability Problems With The Quantum Approximate Optimization Algorithm},
  author = {Boulebnane, Sami and Montanaro, Ashley},
  journal = {PRX Quantum},
  volume = {5},
  issue = {3},
  pages = {030348},
  numpages = {32},
  year = {2024},
  month = {Sep},
  publisher = {American Physical Society},
  doi = {10.1103/PRXQuantum.5.030348},
}

@article{
doi:10.1126/sciadv.adm6761,
author = {Ruslan Shaydulin  and Changhao Li  and Shouvanik Chakrabarti  and Matthew DeCross  and Dylan Herman  and Niraj Kumar  and Jeffrey Larson  and Danylo Lykov  and Pierre Minssen  and Yue Sun  and Yuri Alexeev  and Joan M. Dreiling  and John P. Gaebler  and Thomas M. Gatterman  and Justin A. Gerber  and Kevin Gilmore  and Dan Gresh  and Nathan Hewitt  and Chandler V. Horst  and Shaohan Hu  and Jacob Johansen  and Mitchell Matheny  and Tanner Mengle  and Michael Mills  and Steven A. Moses  and Brian Neyenhuis  and Peter Siegfried  and Romina Yalovetzky  and Marco Pistoia },
title = {Evidence of scaling advantage for the quantum approximate optimization algorithm on a classically intractable problem},
journal = {Science Advances},
volume = {10},
number = {22},
pages = {eadm6761},
year = {2024},
doi = {10.1126/sciadv.adm6761},
}

@article{Tate2023warmstartedqaoa,
  doi = {10.22331/q-2023-09-26-1121},
  title = {Warm-{S}tarted {QAOA} with {C}ustom {M}ixers {P}rovably {C}onverges and {C}omputationally {B}eats {G}oemans-{W}illiamson's {M}ax-{C}ut at {L}ow {C}ircuit {D}epths},
  author = {Tate, Reuben and Moondra, Jai and Gard, Bryan and Mohler, Greg and Gupta, Swati},
  journal = {{Quantum}},
  issn = {2521-327X},
  publisher = {{Verein zur F{\"{o}}rderung des Open Access Publizierens in den Quantenwissenschaften}},
  volume = {7},
  pages = {1121},
  month = sep,
  year = {2023}
}

@misc{farhi2019quantumsupremacyquantumapproximate,
      title={Quantum Supremacy through the Quantum Approximate Optimization Algorithm}, 
      year ={2016},
      author={Edward Farhi and Aram W Harrow},
      eprint={1602.07674},
      archivePrefix={arXiv},
      primaryClass={quant-ph},
      doi={10.48550/arXiv.1602.07674}
}

@article{Born1928,
author = {Born, M and Fock, V},
doi = {10.1007/BF01343193},
issn = {0044-3328},
journal = {Zeitschrift f{\"{u}}r Phys.},
number = {3},
pages = {165--180},
title = {Beweis des Adiabatensatzes},
volume = {51},
year = {1928}
}

@INPROCEEDINGS{10313741,
  author={Golden, John and Bärtschi, Andreas and O'Malley, Daniel and Eidenbenz, Stephan},
  booktitle={2023 IEEE International Conference on Quantum Computing and Engineering (QCE)}, 
  title={Numerical Evidence for Exponential Speed-Up of QAOA over Unstructured Search for Approximate Constrained Optimization}, 
  year={2023},
  volume={01},
  number={},
  pages={496-505},
  doi={10.1109/QCE57702.2023.00063}}

@article{Wilson2021,
author = {Wilson, Max and Stromswold, Rachel and Wudarski, Filip and Hadfield, Stuart and Tubman, Norm M and Rieffel, Eleanor G},
doi = {10.1007/s42484-020-00022-w},
issn = {2524-4914},
journal = {Quantum Mach. Intell.},
number = {1},
pages = {13},
title = {{Optimizing quantum heuristics with meta-learning}},
volume = {3},
year = {2021}
}

@article{PhysRevA.98.032309,
  title = {Quantum circuit learning},
  author = {Mitarai, K. and Negoro, M. and Kitagawa, M. and Fujii, K.},
  journal = {Phys. Rev. A},
  volume = {98},
  issue = {3},
  pages = {032309},
  numpages = {6},
  year = {2018},
  month = {Sep},
  publisher = {American Physical Society},
  doi = {10.1103/PhysRevA.98.032309},
}

@INPROCEEDINGS{bucher2024robustbenchmarkingquantumoptimization,
  author={Bucher, David and Kraus, Nico and Blenninger, Jonas and Lachner, Michael and Stein, Jonas and Linnhoff-Popien, Claudia},
  booktitle={2024 IEEE International Conference on Quantum Computing and Engineering (QCE)}, 
  title={Towards Robust Benchmarking of Quantum Optimization Algorithms}, 
  year={2024},
  volume={03},
  number={},
  pages={159-170},
  doi={10.1109/QCE60285.2024.11030870}}

@inproceedings{10.1145/3583133.3596358,
author = {Stein, Jonas and Chamanian, Farbod and Zorn, Maximilian and N\"{u}\ss{}lein, Jonas and Zielinski, Sebastian and K\"{o}lle, Michael and Linnhoff-Popien, Claudia},
title = {Evidence that PUBO outperforms QUBO when solving continuous optimization problems with the QAOA},
year = {2023},
isbn = {9798400701207},
publisher = {Association for Computing Machinery},
address = {New York, NY, USA},
doi = {10.1145/3583133.3596358},
booktitle = {Proceedings of the Companion Conference on Genetic and Evolutionary Computation},
pages = {2254–2262},
numpages = {9},
location = {Lisbon, Portugal},
series = {GECCO '23 Companion}
}

\end{document}